\documentclass[nonacm,sigplan]{acmart}

\usepackage[]{hyperref}

\usepackage{grumble}
\usepackage{braket}

\newcommand{\projecttitle}{QVM}

\newcommand{\myparagraph}[1]{ \noindent{\bf {#1}.}}

\newcommand{\out}[1] {}

\newcounter{codeLineCntr}

\reversemarginpar
\newif\ifnotes
\notestrue

\newcommand{\punt}[1]{}

\renewcommand{\eqref}[1]{Equation~(\ref{eq:#1})}

\newcommand{\proc}[1]{\ifmmode\mbox{\textsc{#1}}\else\textsc{#1}\fi}

  \newcommand{\func}[1]{\ifmmode\mathrm{#1}\else\textrm{#1}fi} %
\newcounter{remark}[section]

\usepackage[inline]{enumitem}
\setlist{noitemsep,topsep=0pt,parsep=0pt,partopsep=0pt}

\usepackage[subtle]{savetrees}

\usepackage{setspace}
\usepackage[margin=5pt,font={stretch=0.9}]{caption}

\begin{document}

\title{Scaling Quantum Computations via Gate Virtualization}

\author{Nathaniel Tornow}
\email{nathaniel.tornow@tum.de}
\affiliation{%
  \institution{TU Munich and Leibniz Supercomputing Centre}
  % \city{Munich}
  \country{Germany}
}
\author{Emmanouil Giortamis}
\email{emmanouil.giortamis@tum.de}
\affiliation{%
  \institution{TU Munich}
  % \city{Munich}
  \country{Germany}
}
% \author{Martin Ruefenacht}
% \email{martin.ruefenacht@lrz.de}
% % \orcid{1234-5678-9012}
% % \author{G.K.M. Tobin}
% % \email{webmaster@marysville-ohio.com}
% \affiliation{%
%   \institution{Leibniz Supercomputing Centre}
%   % \city{Munich}
%   \country{Germany}
% }
\author{Pramod Bhatotia}
\email{pramod.bhatotia@tum.de}
\affiliation{%
  \institution{TU Munich}
  % \city{Munich}
  \country{Germany}
}

\begin{abstract}

We present the Quantum Virtual Machine (QVM), an end-to-end generic system for scalable execution of large quantum circuits with high fidelity on noisy and small quantum processors (QPUs) by leveraging gate virtualization.
\projecttitle{} exposes a virtual circuit intermediate representation (IR) that extends the notion of quantum circuits to incorporate gate virtualization. Based on the virtual circuit as our IR, we propose the \projecttitle{} compiler---an extensible compiler infrastructure to transpile a virtual circuit through a series of modular optimization passes to produce a set of optimized circuit fragments. Lastly, these transpiled circuit fragments are executed on QPUs using our \projecttitle{} runtime---a scalable and distributed infrastructure to virtualize and execute circuit fragments on a set of distributed QPUs.

We evaluate \projecttitle{} on IBM's 7- and 27-qubit QPUs.
Our evaluation shows that using our system, we can scale the circuit sizes executable on QPUs up to double the size of the QPU while improving fidelity by 4.7$\times$ on average compared to larger QPUs and that we can effectively reduce circuit depths to only 40\% of the original circuit depths.
%Finally, we show that QVM can scale circuit simulation up to 100 qubits, outperforming classical simulators. 
% \pramod{update me!}
\end{abstract}

\maketitle % should come after the abstract
\pagestyle{plain} % should come right after \maketitle

% \end{abstract}

\section{Introduction}
Quantum computers promise to solve otherwise intractable problems in optimization \cite{farhi2014quantum}, factorization \cite{shor1999polynomial}, or quantum simulation \cite{kandala2017hardware, peruzzo2014variational}.
However, the reliable operation of quantum processing units (QPUs) is extremely challenging, as the same properties that could lead to computational benefits are also the main reason for uncontrollable noise and state-decoherence during a quantum computation on a QPU\cite{preskill2018quantum}.
This still severely limits the number of qubits and operations we can run within the same quantum program.
% by exploiting the laws of quantum mechanics.
% Recent advancements led to the development of a range of quantum processing units (QPUs) with hundreds of qubits \cite{ibmQuantum, quantumaiHardwareGoogle, ionqIonQTrapped}. However, with quantum error correction still being well out of reach, today's QPUs are still considered Noisy Intermediate Scale Quantum (NISQ) QPUs, characterized by low qubit numbers, noisy operations, and low coherence times \cite{preskill2018quantum}.
% With these small and noisy QPUs, we are, therefore, severely limited in the number of operations and the number of qubits on which programs can run with high fidelity.

\textit{Gate virtualization} (GV) has recently been proposed to scale the size of quantum programs running with high fidelity on small and noisy QPUs \cite{mitarai2021constructing}.
This technique virtualizes binary qubit gates by sampling the gate with single qubit operations instead.
Theoretical work shows that GV allows quantum circuits to be optimized to scale and improve fidelity in two different dimensions: First, quantum circuits can be decomposed into multiple smaller circuit fragments to run on small QPUs \cite{mitarai2021constructing, peng2020simulating, piveteau2022circuit}, and second, circuit depth can be reduced to increase overall fidelity \cite{circuit-knitting-toolbox, yamamoto2022error}.
% Thus, when virtualizing binary gates in a quantum circuit, we could benefit from two main opportunities:
% First, when virtualizing binary gates, we can remove the dependence between qubits in the circuit. This prevents error propagation in the circuit, allows us to run the circuit with fewer noisy SWAP gates on constrained QPU topologies \cite{}, and enables qubit reuse within the circuit to reduce the number of qubits required.
% Second, if we break the qubit dependencies sufficiently, we can decompose a large circuit into several smaller circuit fragments with fewer qubits and fewer operations, allowing us to simulate large circuits on small NISQ QPUs with high fidelity.
% more weak
% 1. promise,
% 2. what is it
% have been applied in a theory manner
% 
% Limitations of gate virtualization
% and challenges (next paragraph)

However, the effectiveness of gate virtualization is severely hampered by the lack of general and extensible procedures for automatically applying and executing gate virtualization. 
Previous studies have primarily concentrated on utilizing gate virtualization through ad-hoc methods or on an individual application level \cite{circuit-knitting-toolbox, yamamoto2022error, bechtold2023investigating}.
% \pramod{maybe we should state that therefore gate virtualization is applied in an ad hoc on a per-application basis?}
% Moreover, the applicability of gate virtualization suffers greatly from the high computational cost of $\mathcal{O}(6^k)$ for virtualizing $k$ binary gates, both in quantum \revision{circuit overhead and classical postprocessing} 
Moreover, the applicability of gate virtualization suffers greatly from the high computational cost, since virtualizing $k$ two-qubit gates comes with a quantum circuit and classical post-processing overhead of $\mathcal{O}(6^k)$ \cite{mitarai2021constructing}.

To this end, we target the following research question: \textit{How can we design a generic and extensible system that fully utilizes the full potential of GV to scale the size of circuits that can be executed with high fidelity on current QPUs, despite the computational overhead?}

To address this research question, in this work, we present the Quantum Virtual Machine (\projecttitle{}), a system for scalable and reliable execution of quantum circuits on small and noisy QPUs by fully leveraging GV.
We make the following key contributions:

% To answer this research question, we need to address three major challenges:
% First, we need generic and extensible abstractions and interfaces to express the new technique of gate virtualization and fragmented quantum circuits.
% Using this abstraction, we need to design automatic optimization techniques that apply as little gate virtualization as possible to minimize the virtualization overhead while getting the most benefit from improving scalability and fidelity.
% Finally, we need a system that can scale the number of gate virtualizations as much as possible through distributed and parallel processing.

% Problem statement: How do we design a generic system to utilize gate virtualization
{\revision

% To address these challenges, we introduce the Quantum Gate Virtualization Machine (\projecttitle{}), a system for scalable and reliable execution of quantum circuits on NISQ QPUs by leveraging gate virtualization.
% We make the following contributions:

\begin{itemize}
    \item To enable the general and programmable application of GV, and as the basis of our work, we present the \textbf{virtual circuit IR} (VC-IR). The VC-IR extends the quantum circuit abstraction, manages virtual gates, decompositions into several smaller circuit fragments, and data structures to efficiently analyze and apply GV (\S~\ref{sec:abstraction}).
    \item To enable automatic and efficient GV that optimizes a circuit with as little post-processing as possible, we introduce the \textbf{\projecttitle{} Compiler}, an extensible pipeline for converting large quantum circuits into optimized virtual circuits (\S~\ref{sec:compiler}).
     The compiler enables multiple optimization passes that apply GVs based on the VC-IR.
     We present three generic optimization passes for efficient gate virtualization on arbitrary quantum circuits.
     (1) The \textit{circuit cutter} (\S~\ref{sec:cc}) decomposes a circuit into several smaller fragments to run on smaller QPUs, (2) the \textit{dependency reducer} (\S~\ref{sec:cdr}) reduces the dependencies within a circuit to reduce error propagation and the number of SWAP gates, and (3) the \textit{qubit reuser} (\S~\ref{sec:qr}) applies a qubit reuse technique to reduce the width of circuit fragments, enabling a trade-off between GV's overhead and circuit depth.
    \item To enable as many GVs as possible to benefit from its opportunities, we present the \textbf{\projecttitle{} Runtime}, a scalable system that can run a virtual circuit (VC) on a set of QPUs and classical nodes (\S~\ref{sec:runtime}).
     The runtime uses the core component of a \textit{virtualizer} (\S~\ref{sec:virtualizer}) that instantiates fragments of the VC and computes the result of the VC using highly parallel post-processing.
     The quantum circuits are executed in a scalable manner on a distributed set of QPUs using \projecttitle{}'s \textit{QPU manager} (\S~\ref{sec:runtime:scheduler}).
\end{itemize}

}

We implement \projecttitle{} in Python by building on the Qiskit framework \cite{Qiskit}.
For our compiler, in addition to heuristic algorithms, we implement optimal passes using Answer Set Programming following our optimization models \cite{gebser2012answer}.

We evaluate \projecttitle{} on IBM's 7- and 27-qubit QPUs and simulators using various circuits used in popular quantum algorithms \cite{tomesh2022supermarq, quetschlich2022mqt, li2023qasmbench}.
{Based on our analysis, we can execute circuits with up to 2$\times$ the number of qubits of the QPU used while improving fidelity by an average of 4.7$\times$ and up to 33.6$\times$ (\S~\ref{sec:eval:circ_cutter})}. Our intra-circuit dependency reduction techniques reduce the depth of transpiled circuits on average to 64\% of the original circuit and increase fidelity by an average of 1.4$\times$ and up to 5.2$\times$ (\S~\ref{sec:eval:dep_reducer}). Our dependency reducer also enables the reuse of more qubits to reduce the width of circuits with less virtualization overhead (\S~\ref{sec:eval:qr}).
Our \projecttitle{} runtime scales and can be used to run large virtual circuits with low memory requirements (\S~\ref{sec:eval:end-to-end-analysis}-\ref{sec:eval:practical}).
\section{Background and Motivation}

% \subsection{Impact of Circuit Properties in the NISQ Era}
% NISQ QPUs face challenges such as significant noise, limited qubit connectivity, and decoherence within microseconds \cite{ibmQuantum, google-nisq-properties}. Additionally, imperfections in quantum gates and measurement operations introduce computational errors during program execution. For instance, CNOT operations have an error rate of $\sim$1\% \cite{ibmQuantum, google-nisq-properties}.  Limited qubit connectivity necessitates non-local SWAP operations, each introducing three more CNOT operations \cite{gushu2019tackling}. Finally, qubit dependencies \cite{hua2023caqr} amplify noise-propagation between operations and qubits \cite{gonzalez2022error}, and restrict circuit placement on a QPU, leading to more SWAP operations. Systems that optimize circuits by reducing their depth, number of CNOT operations, and qubit dependencies, are essential for practical quantum computing in the NISQ era. 

% State-of-the-art transpilers \cite{molavi2022qubit, gushu2019tackling, qiskit-transpiler, swamit2019not, cowtan2019qubit, sivarajah2020t} mainly focus on minimizing the circuit's post-transpilation depth and number of CNOTs. 
% Approaches that substitute the noisy and connectivity-restricted binary gates with unary gates can be used to improve all key properties of a circuit, including its width and qubit dependencies, mostly at the cost of computational overheads \cite{mitarai2021constructing, peng2020simulating, saleem2021quantum, yamamoto2022error}.

\subsection{Computational Model}

We define a quantum computation by adopting the computational model of \cite{peng2020simulating}.
A quantum computation is defined by a quantum circuit with $m$ qubits and one- and two-qubit gates, initialized with the $\ket{0}^{\otimes m}$ state (Fig.~\ref{fig:formula}, (a)).
All output qubits are measured in the computational basis to obtain a measurement bitstring $x \in \{0,\ 1\}^m$. We then apply a classical post-processing function $f: x \to [-1,\ 1]$ to the bitstrings. We aim to accurately approximate the expectation value $\mathbb{E}_x\ f(x)$ of the function across the bitstrings.

In our work, we focus on measuring expectation values of $m$-qubit projection operators $O = P_y$, i.e., we measure the expectation value $\braket{P_y}$ of the observable $P_y$. In our case, $f$ is therefore defined as $f(x) = 1$ if $x = y$, otherwise $f(x) = 0$.
This is the same as calculating the probability of obtaining a certain output string $y$, i.e., $\braket{P_y} = p(y)$.
$\braket{P_y}$ can, therefore, be determined by sampling the quantum circuit many times and calculating the ratio of the shots that returned $y$ to the total number of shots.

\subsection{Foundations of Gate Virtualization}\label{sec:back:gate-virt}
\begin{figure}[t] 
    \centering 
    \includegraphics[width=.79\columnwidth]{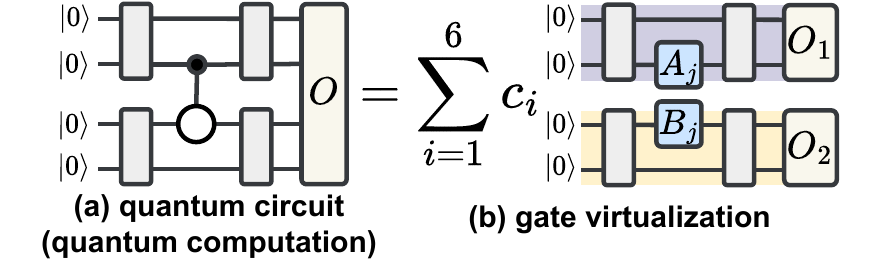} 
    \caption{
    Computational model and gate virtualization (\S~\ref{sec:back:gate-virt}). {\em
    (a) A quantum computation to estimate the expectation value of an observable $O$. (b) Virtualizing a two-qubit gate by computing a weighted sum over circuit instances with single-qubit gates inserted ($O = O_1 \otimes O_2$)
    }} 
    \label{fig:formula}
\end{figure}

Gate virtualization (GV) allows us to decompose two-qubit gates into a combination of single-qubit gates \cite{mitarai2021constructing}.
We show GV schematically in Fig.~\ref{fig:formula} (b). Instead of executing the original circuit's binary gate, we can calculate a weighted sum of six circuit instances to estimate $\braket{O}$. In each circuit instance $i$, instead of the original two-qubit gate, we insert $A_i$ and $B_i$, which are either one-qubit unitary gates or projective measurements. This allows us to decompose each instance $i$ into two smaller, completely independent sub-circuits, which can be sampled independently. We reconstruct the result with
\begin{equation}
    \braket{O} = \sum_{i=1}^6 c_i \braket{O}_i = \sum_{i=1}^6 c_i \braket{O_1}_i \braket{O_2}_i ,
    \label{eq:qpd}
\end{equation}
where $\braket{O}_i$ is the result of each instance $i$ and $O = O_1 \otimes O_2$.
The exact proofs and formulas of GV for standard gates such as CNOT or $R_{ZZ}$ can be found in \cite{mitarai2021constructing}.

\myparagraph{Virtualizing Mulitple Gates}
We now generalize GV to be applied on $k$ gates in a circuit.
We can think of adding a GV of another gate in a circuit as performing an additional GV for each instance of the original virtual gate.

To formalize the QPD of multiple gates in a circuit, let $G_v$ be the set of all virtual gates in a quantum circuit.
We then define a coefficient vector for each virtual gate $g \in G_v$ as $\mathbf{c}_g = (c_1,\, ...,\ c_{6})$. We define the global coefficient vector as
\begin{equation}
    \mathbf{C} = \bigotimes_{g \in G_v}\ \mathbf{c}_g,
    \label{eq:coeff}
\end{equation}
i.e., the tensor product of all individual coefficient vectors $c_g$.
Therefore, $\mathbf{C}$ is a vector with $|\mathbf{C}| = 6^k$ entries.
To reconstruct the final results, we calculate
\begin{equation}
    \braket{O} = \sum_{c_i \in \mathbf{C}} c_i\ \prod_{j=1}^f \braket{O_j}_i ,
    \label{eq:general_qpd}
\end{equation}
which is the generalization of Eq.~(\ref{eq:qpd}).
Here $f$ is the number of subcircuits, and $\braket{O_j}_i$ is the result of the $j$th subcircuit in the $i$th global instance.
We must, therefore, calculate a sum over $|\mathbf{C}| = 6^k$ elements. Since in general $|\mathbf{C}| \gg f > |G_v|$, we need $\mathcal{O}(6^k)$ operations to calculate Eq.~(\ref{eq:general_qpd}).

GV thus causes an exponential post-processing overhead of $\mathcal{O}(6^k)$ \cite{mitarai2021constructing}. This severely limits the number of gates that can be virtualized within a circuit, meaning that we need to find a good compromise between the additional runtime and the benefits of GV, as described in the next section.

\subsection{Opportunities of Gate Virtualization}
State-of-the-art circuit transpilers \cite{qiskit-transpiler} mainly focus on minimizing the circuit's post-transpilation depth and number of CNOTs. 
GV can be used effectively to reduce the width and qubit dependencies in a circuit, leading to improved execution fidelity for larger circuits on small QPUs, mostly at the cost of computational overheads, as shown in previous ad-hoc and theoretical work \cite{mitarai2021constructing, peng2020simulating, saleem2021quantum, yamamoto2022error}.
In total, GV gives us opportunities in the following two dimensions:

\myparagraph{Cutting Quantum Circuits}
By virtualizing binary gates, a circuit can be divided into smaller subcircuits, each with a lower number of qubits which can be run independently on small and noisy QPUs \cite{mitarai2021constructing, peng2020simulating}. Circuits with lower widths exhibit less qubit mapping and routing constraints, therefore less post-transpilation CNOT operations and lower depth \cite{nation2023suppressing}. Moreover, as recent work shows, the wire-cutting technique \cite{peng2020simulating, tang2021cutqc} can also be modeled using gate virtualization \cite{brenner2023optimal}.
Therefore, a system that is able to virtualize gates is a complete solution for circuit cutting and knitting.

\myparagraph{Reducing Qubit Dependencies}
Gate virtualization can be used to virtualize binary gates that cause qubit dependencies. Virtualizing them and therefore reducing qubit dependencies in a circuit has three-fold advantages:
Firstly, by reducing the propagation of errors through gates and qubits, fidelity can be improved.
Secondly, reducing intra-circuit dependencies facilitates optimized qubit mapping and routing on QPUs during the transpilation process, which leads to lower depth and number of CNOTs.
Lastly, fewer qubit dependencies enable the application of qubit-reuse \cite{hua2023caqr}, which could enable a computationally efficient way to reduce circuit width.

To summarize, GV is a promising technique to improve the execution accuracy and scaling of quantum circuits, as shown in small ad-hoc examples in previous work.
However, to fully exploit the benefits of GV on arbitrary circuits despite the exponential post-processing overhead, we need an automatic and efficient application as well as a scalable execution of GV.

\section{Overview}

We aim to design a system that can scale the sizes of circuits that can practically be executed on QPUs with high fidelity. To realize this goal, gate virtualization (GV) is a promising technique that decomposes circuits into smaller circuit fragments or reduces the intra-dependencies in the circuit. However, dealing with the programming and exponential computational complexity of GV is challenging. We next discuss these challenges and present our key ideas to address them.

\subsection{Design Challenges and Key Ideas}

\myparagraph{Challenge \#1: Programmability and Generality} 
The promising technique of GV is a new and rather complex concept. It is not trivial how to virtualize two-qubit gates using single-qubit gates or how to keep track of the created circuit fragments.
Therefore, we must develop general abstractions that implement the new virtualization techniques and allow simple, automatic application to quantum circuits while allowing straightforward integration into existing transpilation and optimization infrastructures. 

\noindent\fbox{\parbox{\columnwidth}{
{\bf Virtual Circuit IR (VC-IR) as an Intermediate Representation:}
We introduce the \textit{virtual circuit IR} (VC-IR)  to enable a unified optimization and execution process of large circuits using gate virtualization. The VC-IR is an intermediate step between any high-level circuit representation and smaller optimized circuit fragments.
}}

\myparagraph{Challenge \#2: Fidelity}
The noisy circuit executions on QPUs hinder the practicality of current quantum algorithms.
Every operation applied on a qubit incurs noise to the final result, which propagates and amplifies throughout the circuit.
To ensure higher fidelity in quantum computations, it is essential to employ procedures that optimize the circuit's structure using the promising technique of gate virtualization. This involves decomposing the circuit into smaller fragments, reducing the circuit's depth, and minimizing the number of non-local operations or qubit dependencies while minimizing the overhead of virtualization.

\noindent\fbox{\parbox{\columnwidth}{
{\bf A Compiler for Optimal Gate Virtualization:}
We introduce the \textit{\projecttitle{} Compiler}, a modular architecture designed to compile circuits utilizing gate virtualization.
The compiler converts a quantum circuit into a VC, applies a customizable series of optimization passes on the VC to take advantage of gate virtualization opportunities, and prepares the VC for execution on a distributed set of QPUs.
% Users can modify the optimization passes to suit their specific requirements.
}}

\myparagraph{Challenge \#3: Scalability}
Gate virtualization incurs an exponential overhead of $\mathcal{O}(6^k)$ for $k$ virtual gates, both in quantum computation and in classical postprocessing (Fig.~\ref{fig:formula}).
This overhead appears since we need to execute the fragments in $\mathcal{O}(6^k)$ instantiations, and then we need to post-process all instantiation results to compute the final result.
To maximize the possible gate virtualizations, it is crucial to implement highly parallel computation on multiple QPUs and classical processors.

\noindent\fbox{\parbox{\columnwidth}{
{\bf A Distributed Scalable Runtime:}
We present \textit{\projecttitle{} Runtime}, a scalable system for executing virtual circuits.
The runtime efficiently instantiates the high amount of fragments, distributes them between available QPUs for parallel quantum processing, and uses a highly scalable parallel process to post-process the fragment results.
}}

\begin{figure}[t]
\centering
  \includegraphics[width=.8\columnwidth]{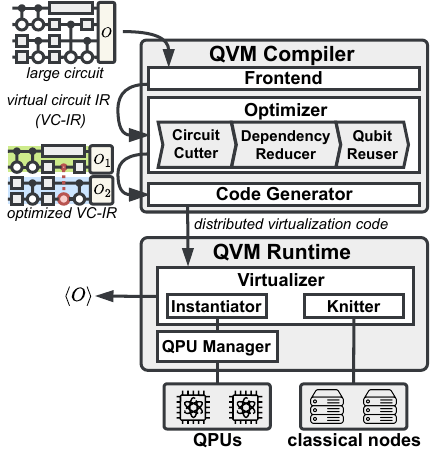}
  \caption{Overview of the \projecttitle{} framework (\S~\ref{sec:framework}). {\em The Quantum Virtual Machine (\projecttitle{})  consists of two main components: QVM Compiler and  QVM Runtime}.
  }
  \label{fig:overview}
\end{figure}

\subsection{The \projecttitle{} Framework}
\label{sec:framework}

Based on the aforementioned key ideas, we propose the design of our Quantum Virtual Machine (\projecttitle{}) framework, an end-to-end system that exploits the full potential of gate virtualization to achieve scalable execution of large circuit with high fidelity (see Fig.~\ref{fig:overview}).  The \projecttitle{} system builds on the abstraction of a \textit{virtual circuit} to utilize gate virtualization. It consists of two main components: the \projecttitle{} Compiler ($\S$~\ref{sec:compiler}) and the \projecttitle{}  Runtime ($\S$~\ref{sec:runtime}).

\myparagraph{\projecttitle{}  Virtual Circuit IR (VC-IR)}
The virtual circuit (VC) abstraction extends the traditional quantum circuit abstraction (\S~\ref{sec:abstraction}). For this, it incorporates the abstraction of \textit{virtual gates} and views the circuit as a collection of circuit \textit{fragments}, where each fragment is a circuit acting on a subset of qubits of the original circuit.

\myparagraph{\projecttitle{} Compiler}
The \projecttitle{} compiler (Fig.~\ref{fig:overview}, top) is responsible for compiling a quantum circuit efficiently to a set of smaller circuit fragments by using gate virtualization.
The compiler operates in three stages: (1) the \textit{frontend} converts the circuit into the VC-IR, (2) the \textit{virtual circuit optimizer} applies gate virtualization to reduce circuit depth, width, and/or intra-circuit dependencies, and (3) the \textit{distributed transpiler} prepares the circuit fragments for execution on a set of QPUs.
For the virtual circuit optimizer, we describe the implementation of three optimization passes of the \textit{circuit cutter}, the \textit{dependency reducer}, and the \textit{qubit reuser}, {\revision which are designed to optimize arbitrary virtual circuits. Additionally, users can easily plug in their own optimization passes, which modify a virtual circuit, e.g., to efficiently optimize specific circuits of a known structure.}

\myparagraph{\projecttitle{} Runtime}
The \projecttitle{} runtime (Fig.~\ref{fig:overview}, bottom) is the system responsible for the scalable execution of virtual circuits. The runtime consists of the two components \textit{virtualizer} and the \textit{scheduler}.
The virtualizer is responsible for implementing the gate virtualizations according to Fig.~\ref{fig:formula}, using fragment instantiation and parallel post-processing.
The QPU manager is responsible for the parallel execution of the fragments on a set of distributed QPUs.

\section{The \projecttitle{} Compiler}
\label{sec:compiler}

% \begin{Fig.*}[t]
%     \centering
%     \includegraphics[width=.85\textwidth]{Fig.s/compiler.png}
%     \caption{Overview of the \projecttitle{} Compiler (\S~\ref{sec:compiler}). {\em The \projecttitle{} Compiler consists of three stages: the Transformer (virtual circuit generation), Optimizer (a modular compiler optimization workflow), and Distributed Transpiler (transpilation for target QPUs).}}
%     \label{fig:comp_overview}
% \end{Fig.*}

\begin{figure}[t]
    \centering
    \includegraphics[width=.95\columnwidth]{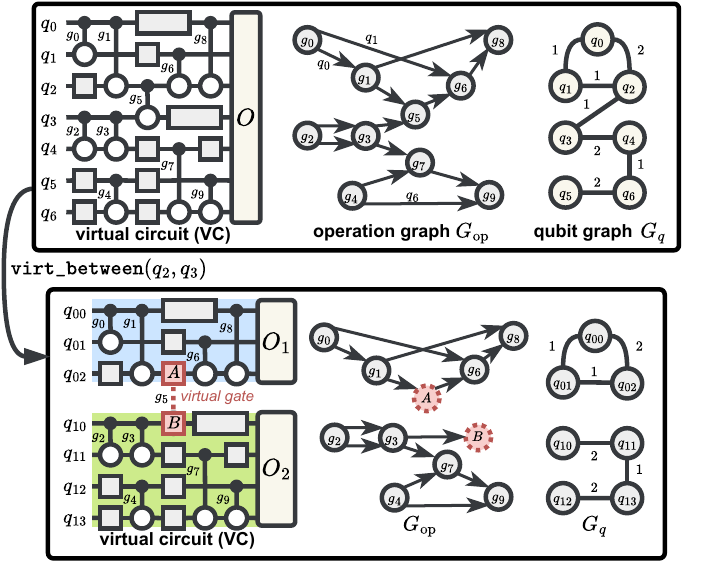}
    \caption{
    \revision
    Virtual Circuit IR (VC-IR) (\S~\ref{sec:abstraction}). 
    {\em A virtual circuit (VC) extends a quantum circuit by incorporating virtual gates and managing the qubits in sets of fragments. The VC-IR manages the operation graph $G_{op}$ and the qubit graph $G_q$ for efficient analysis and manipulation with the VC-IR API to apply GV.
    }}
    \label{fig:virtual_circuit}
\end{figure}

We now describe the design and implementation of our \projecttitle{} compiler.
The \projecttitle{} compiler is an extensible pipeline for the efficient virtualization of gates and to prepare a large circuit for executing a set of small QPUs.
% The \projecttitle{} compiler is responsible for transforming a circuit into an optimized virtual circuit (VC) and preparing its fragments for execution on a distributed set of NISQ QPUs.

\subsection{Workflow of the \projecttitle{} Compiler}
Fig.~\ref{fig:overview} (top) shows the workflow of the \projecttitle{} compiler. First, the \textbf{frontend} of our compiler takes a (large) quantum circuit and converts the circuit into the virtual circuit IR (VC-IR) (Fig.~\ref{fig:virtual_circuit}).

Then, the VC-IR is optimized using the \textbf{optimizer}. Each compiler optimization pass receives two inputs: the maximum fragment size $s$, which specifies the maximum width each fragment must have, and a virtualization budget $b$, which constrains the number of allowed gate virtualizations to limit the maximum virtualization overhead. Typically, we choose $s$ as the size of the largest available QPU to ensure every fragment is executable by at least one QPU.
For our optimizer, we design a pipeline of the following three generic optimization passes:

\noindent
{\bf \#1: Circuit Cutter (CC)}:
First, the circuit cutter (CC) pass (\S~\ref{sec:cc}) aims to decompose the VC into fragments smaller than $s$, using $v \leq b$ virtual gates, and sets the budget to $b = b - v$. If CC fails to decompose the circuit within the given budget, no gate is virtualized and we forward the VC to the next stage.

\noindent
{\bf \#2: Dependency Reducer (DR)}:
In the case where the budget $b$ is not yet exhausted by the circuit cutter pass, the dependency reducer (DR) (\S~\ref{sec:cdr}) applies up to $b$ gate virtualizations to reduce the dependencies between qubits and operations within the VC to reduce noise propagation and circuit depth.

\noindent
{\bf \#3: Qubit Reuser (QR)}:
Lastly, the qubit reuser (\S~\ref{sec:qr}) reuses qubits within individual fragments to further reduce circuit width if the CC fails to reduce the fragment sizes sufficiently.
% The previous DR pass enables this qubit reuse and does not affect the gate virtualization further since it only operates on individual circuit fragments.
If the qubit reuser fails to reduce the width of any fragment to $s$, the optimization pipeline fails.

After the optimization phase, the \textbf{code generator} (CG) acts as the backend of our compiler (\S \ref{sec:DT}) by extracting the fragments as parameterized circuits, optimizing the circuits and generating the inputs for the instantiation of each fragment.

Next, we describe the \projecttitle{} compiler stages in detail.

\begin{figure}[t]
    \centering
    \includegraphics[width=.95\columnwidth]{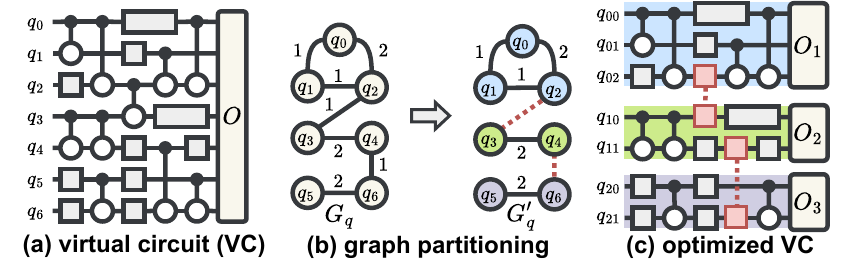}
    \caption{Circuit Cutter (CC) (\S~\ref{sec:cc}). {\em The CC receives a large virtual circuit (VC) with $n_q=6$ qubits and performs a graph partitioning on the qubit graph $G_q$ to dissect the VC into fragments of size $s=3$ by inserting virtual gates between the partitions.}}
    \label{fig:gate_decomp}
\end{figure}

\begin{figure*}[t]
    \centering
        \includegraphics[width=.96\textwidth]{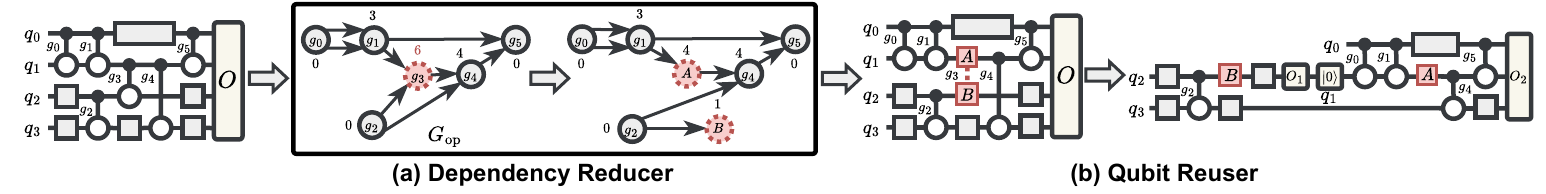}
    \caption{Dependency Reducer (DR) and Qubit Reuser (QR) (\S~\ref{sec:cdr}-\ref{sec:qr}). {\em (a) The greedy DR iteratively virtualizes gates to reduce the number of qubit-dependencies in a circuit. (b) Because of reduced qubit dependencies, the QR can reuse qubits to reduce the circuit width.}}
    \vspace{-5mm}
    \label{fig:dr}
\end{figure*}

\subsection{Virtual Circuit IR and Frontend} 
\label{sec:abstraction}

To enable easy integration and a simple workflow for gate virtualization during compilation and runtime, \projecttitle{} provides the virtual circuit IR (VC-IR) (Fig.~\ref{fig:virtual_circuit}).
In total, the VC-IR provides three main data structures: (1) A \textit{virtual circuit} (VC), which can contain \textit{virtual gates} and consists of several \textit{fragments}, (2) an operation graph $G_{op}$ and (3) a qubit graph $G_q$, as described below:

\myparagraph{Virtual Circuit and Virtual Gates}
A VC extends the traditional abstraction of a quantum circuit by additionally incorporating the functionality of consisting of a set of fragments and allowing \textit{virtual gates} to be part of its instructions.

A \textit{fragment} describes a subset of qubits that are not connected to other qubits in the VC via a real two-qubit gate.
We implement fragments by using a separate qubit register for each fragment.

A \textit{virtual gate} expresses the notion of the virtualization of a binary quantum gate (\S~\ref{sec:back:gate-virt}). A virtual gate is a binary quantum gate that does not require a real connection between its two qubits. Therefore, a conventional transpiler or circuit optimizer would treat a virtual gate as two one-qubit gates. Hence, a virtual gate has no influence on, e.g., the assignment and routing of qubits.
A virtual gate can be split into two parameterized one-qubit gates, whose instantiations are inserted during execution (\S~\ref{sec:virtualizer}).

% Since the VC merely extends the concept of a quantum circuit, any conventional circuit optimization or transpilation also works for the VC without any changes.

\myparagraph{Operation Graph}
The operation graph $G_{op}$ expresses the gate dependencies of the VC as a directed acyclic graph (DAG).
$G_{op}$ is a graph in which the vertices are the two-qubit gates of the circuits, and the edges represent the direct dependencies between the respective operations via a qubit wire. Therefore, each edge contains the respective qubit as an attribute.

\myparagraph{Qubit Graph}
To efficiently represent the connections between qubits of a VC, we utilize the representation of a qubit graph $G_q$, where the qubits are the vertices. An edge exists between two qubits when the qubits are connected with at least one two-qubit gate.
So, the connected subgraphs of $G_q$ directly correspond to the VC's fragments.
Each edge holds a weight with the number of two-qubit gates between the two qubits.

\myparagraph{Gate Virtualization API}
To efficiently virtualize gates, the VC-IR exposes two main functions:
\begin{itemize}
    \item \texttt{virt\_gate}($g_x$): Virtualizes the gate $g_x$, removes $g_x$ from $G_{op}$ and adds single-qubit gates instead. Decrements the weight on the edge ($q_i$, $q_j$) in $G_q$, where $g_x$ acts on $q_i$ and $q_j$.
    \item \texttt{virt\_between}($q_i$, $q_j$): Virtualizes every gate which acts on the qubits $q_i$ and $q_j$. Removes the edge ($q_i$, $q_j$) from $G_q$, and updates $G_{op}$ accordingly.
\end{itemize}
Fig.~\ref{fig:virtual_circuit} shows an example of calling \texttt{virt\_between}($q_i$, $q_j$).

\myparagraph{Frontend: Virtual Circuit Generation}
The frontend of the \projecttitle{} compiler generates the VC-IR from an input circuit. The VC is initially a copy of the original circuit, i.e. a VC that consists of one fragment and no virtual gates.
We generate $G_{op}$ and $G_q$ by traversing the operations of the original circuit.

% \myparagraph{Fragment Extraction}
% A virtual circuit consists of fragments, each representing a distinct subset of qubits in a quantum circuit that is not connected to other fragments by multi-qubit gates, except virtual gates.
% By replacing binary gates with their respective virtual gate during a virtual circuit optimization pass, fragments are automatically restructured in the virtual circuit, in case the insertion of a virtual gate causes breaking the dependency between two qubit-subsets.

% \begin{Fig.}[t]
%     \centering
%     \includegraphics[width=.9\columnwidth]{Fig.s/dr3.png}
%     \caption{Qubit Reuser (\S~\ref{sec:qr}). {\em The width of a virtual circuit can be further reduced by reusing qubit $q_0$ for $q_3$ through measurement and reset. This is only possible because gate $g_2$ was virtualized by the dependency reducer (Fig. \ref{fig:dr}), and therefore $q_3$ does not depend on $q_0$.}} 
%     \label{fig:dr-qr}
% \end{Fig.}

\subsection{Circuit Cutter (CC)}
\label{sec:cc}
The aim of the Circuit Cutter (CC) optimization pass is to split the VC into several fragments so that each fragment has $s$ or fewer qubits, while using as minimal virtual gates as possible to minimize the computational overhead (Fig. \ref{fig:gate_decomp}).
% It aims to minimize the number of virtual gates to reduce the induced overhead while also keeping the circuit fragments as equally sized as possible.
For this purpose, the CC performs a graph partitioning on the qubit-graph $G_q$ as follows:

\myparagraph{Circuit Cutter Model}
We assign the vertices $q_x \in V_q$ of the qubit graph $G_q = (V_q,\ E_q)$ into at least $f = \lceil n_q / s \rceil$ subsets $F_j$. According to this mapping, $E_{cut} = \{ (q_x, q_y)\ :\ q_x \in F_j,\ q_y \in F_i,\ F_j \neq F_i \}$ is the set of all edges that need to be removed to decompose the $G_q$ into independent subgraphs.
In our cutting model, we find a solution that minimizes $\sum_{(q_x, q_y) \in E_{cut}} w(q_x, q_y)$, where $w(q_x, q_y)$ is the weight of the respective edge $(q_x, q_y) \in E_{cut}$.
Amongst all possible optimal solutions that amount to the weight, we choose a solution that minimizes $\sum_j\ |F_j|^2$, such that we favor the solution that distributes the number of qubits evenly across the fragments. The subsets $F_j$ correspond to fragments of the resulting optimized VC.

We implement the model with Answer Set Programming (ASP) using the Clingo solver to find an optimal solution \cite{gebser2012answer, clingo}.
For each $(q_x, q_y) \in E_{cut}$ we call \texttt{virt\_between}($q_x$, $q_y$) to update the VC-IR according to the solution of the model.

\myparagraph{Greedy Circuit Cutter}
In addition to the procedure that finds an optimal solution for our model, we also implement a CC that uses an efficient heuristic approach based on the greedy Kernigan-Lin bisection algorithm \cite{kernighan1970efficient} to enable shorter compilation times for large circuits.
To decompose a VC into multiple fragments of size $s$ or less, we iteratively apply the Kernighan-Lin bisection to the currently largest connected subgraph of $G_q$. The bisection determines two distinct sets of vertices $V_1$ and $V_2$ such that $|V_1| \approx |V_2|$, and the sum of weights of the edges between the two sets of vertices is as minimal as possible.
Then we call \texttt{virt\_between}($q_x$, $q_y$) for each ($q_x$, $q_y$), where $q_x \in V_1$ and $q_y \in V_2$.
We apply this iteration until each fragment of the VC has less than $s$ qubits.

Note that in the partitioning algorithms for gate virtualization, the search space scales only one-dimensionally with the number of qubits in the circuit and not also with the number of binary gates in the circuit, as is the case with wire-cutting \cite{tang2021cutqc}.
Since the number of binary gates in a circuit is typically much larger than the number of qubits, our gate-cutting techniques are generally much more efficient than their wire-cutting counterparts.
This makes optimal graph partitioning for gate virtualization a suitable option for current quantum algorithms with hundreds of qubits and few partitions, where the circuit cutting time is negligible compared to execution time.

\subsection{Dependency Reducer (DR)}
\label{sec:cdr}

The Dependency Reducer (DR) reduces the number of circuit intra-dependencies circuit using as few virtual gates as possible

The dependencies between qubits and operations are best illustrated with the VC-IR's operation graph $G_{op}$.
An example of a $G_{op}$ in the context of qubit dependencies is shown in Fig. \ref{fig:dr} (a).
In this example, every qubit $q_i$ is dependent on every other qubit $q_j$, since some gate $g_x$ acting on $q_i$ depends on a gate $g_y$ acting on $q_j$ \cite{hua2023caqr}.
This means that noise occurring on one qubit could also propagate to all other qubits in the circuit, amplifying overall errors.

As shown in Fig. \ref{fig:dr}, the DR can reduce the intra-dependency of the circuit by inserting virtual gates into the circuit while being constrained by the budget $b$ of the maximum gate virtualizations.
To reduce the qubit-dependencies as efficiently as possible, we adhere to the following model:

\myparagraph{Dependency Reducer Model}
We aim to minimize the number of qubit-dependencies by virtualizing gates in the VC-IR.
A qubit $q_i$ is dependent on another qubit $q_j$ if there exists a path in $G_{op}$ from a gate $g_x$ acting on $q_j$ to a gate $g_y$ acting on $q_i$.
Let $D_q = \{ (q_i, q_j) : q_i \text{ depends on } q_j \}$ be the set of all qubit-dependencies.
We need to find a set $G_{virt}$ of gates that, when removed from $G_{op}$, minimize the number of qubit dependencies $|D_q|$ the most.
If multiple optimal solutions exist, we choose a solution that minimizes $|G_{virt}|$.
We implement this model using Answer Set Programming (ASP) and use the Clingo solver to solve for an optimal solution \cite{gebser2012answer, clingo}.

\begin{figure}
    \centering
    \includegraphics[width=.9\columnwidth]{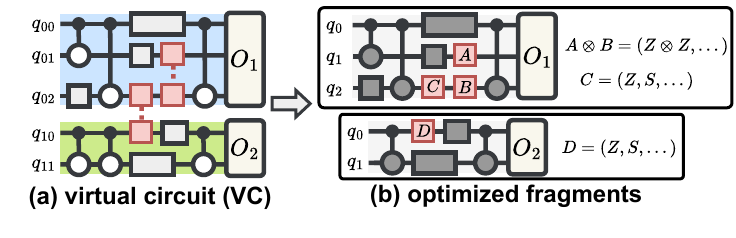}
    \caption{Code Generator (CG) (\S~\ref{sec:DT}). {\em The CG generates distributed virtualization code in form of parameterized fragments.}}
    \label{fig:code-gen}
\end{figure}

\begin{figure*}[t]
    \centering 
    \includegraphics[width=.85\textwidth]{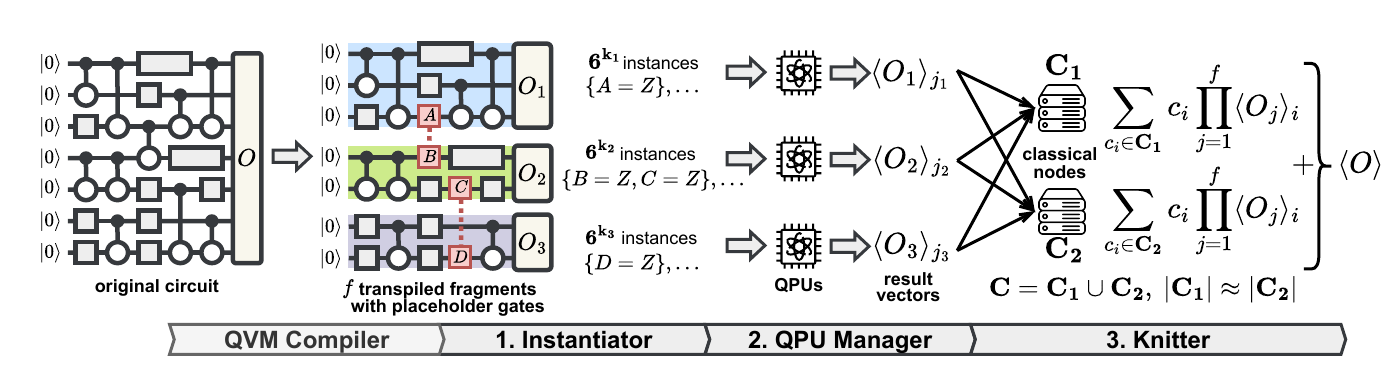} 
    \caption{
    \revision
    Workflow of the QVM Runtime (\S~\ref{sec:runtime}). {\em (1) The instantiator generates the instantiations inserted into the placeholder gates of the compiled fragments. (2) The QPU manager runs the instances on distributed QPUs in parallel. (3) The knitter reconstructs the probability distribution of the original circuit by merging and then knitting the instances in highly parallelizable steps.
    }}
    \vspace{-5mm}
    \label{fig:runtime_overview}
\end{figure*}

\myparagraph{Greedy Dependency Reducer}
For circuits with a large number of gates, we additionally design an efficient greedy DR (G-DR) algorithm (Fig.~\ref{fig:dr}).
The algorithm works as follows:

First, we determine the most critical binary gate in the circuit, i.e., the binary gate that, when virtualized, can reduce the most intra-circuit dependencies. For this, we label every binary gate $g_x$ in the circuit with cost $d_i$. This cost is defined as $d_i = anc(g_x) \cdot desc(g_x)$, where $anc(g_x)$ is the number of ancestors, and $desc(g_x)$ is the number of descendants of $g_x$. Therefore, the gate with the highest cost depends on most other gates in the circuit and is, therefore, most likely responsible for a large amount of qubit and gate dependencies.
Then, we call \texttt{virt\_gate}($g_x$) on the most-critical gate $g_x$ to virtualize the gate in the VC and remove the two-qubit gate from $G_{op}$. If two or more gates have the same cost, we choose a gate randomly.
We decrement the budget $b$, and repeat the process until $b = 0$.

In the example from Fig.~\ref{fig:dr} (a), the G-DR would virtualize $g_3$ first, since it has the highest single cost of $d_3 = anc(g_3) \cdot desc(g_3) = 3 \cdot 2 = 6$.
In this single iteration of G-DR, we can reduce the number of qubit dependencies from 12 to 11 since now $q_2$ does not depend on $q_1$ anymore.
This means that errors of $q_1$ cannot propagate to $q_2$.
It also reduces the cost of all the gates in the circuit, meaning that the gates depend on significantly fewer other gates and are, therefore, less likely to amplify overall noise.

Note that our G-DR computes the number of ancestors for each node in a single traversal of $G_g$ in topological order.
Similarly, the number of descendants of each node is computed in a single traversal in reversed order.
Therefore, the time complexity of G-DR is $\mathcal{O}(2 \cdot n_v \cdot |V_{g}|)$, where $|V_{g}|$ is the set of nodes in $G_g$. Thus, the algorithm has linear time complexity in the number of gates in the circuit.

% \myparagraph{Optimal Dependency Reducer (O-DR)}
% To find the optimal set of at most $b$ binary gates to virtualize to minimize the intra-dependencies of a circuit, we implement an optimal DR (O-DR) procedure.
% The O-DR searches over all possible combinations of $0 - b$ binary gates in the circuit. 
% A combination of $0 - b$ binary gates that, when removed, reduces the qubit-dependencies of the circuit the most is then chosen for virtualization. 
% If multiple combinations of different sizes minimize this dependency by the same amount, the O-DR pass chooses the one with the minimal binary gates to virtualize.
% While this approach is guaranteed to find an optimal solution, it is only feasible for small circuits since the search space scales exponentially with $b$ and the number of binary gates in the circuit.
% We implement the O-DR pass by encoding this problem on $G_g$ into an Answer Set Program (ASP) to solve it with the Clingo ASP solver .

\subsection{Qubit Reuser (QR)}
\label{sec:qr}

In the final pass of the optimizer, we apply the qubit reuser on individual fragments to reduce their width further, in case their width still exceeds the maximal size $s$.
To this end, the qubit reuser first checks whether each fragment in the VC has a width of $s$ or less.
For each fragment with a width greater than $s$, the qubit reuser applies a qubit reuse procedure to reduce the width to $s$ to ensure that each fragment can execute on the available QPUs.
We can reuse a qubit $q_i$ for another qubit $q_j$ if $q_i$ does not depend on $q_j$ by inserting a mid-circuit measurement and resetting the qubit \cite{hua2023caqr}.
Fig.~\ref{fig:dr} (b) shows this qubit reuse pass, where we can reuse $q_2$ for $q_1$ since $q_2$ does not depend on $q_1$, with $O = O_1 \otimes O_2$

The appropriate level of qubit reuse may only be possible through the preceding DR pass, which reduces the number of dependent qubit pairs as shown by the example of Fig.~\ref{fig:dr}.
Similar reduction in width by circuit cutting would have required two virtual gates (or two wire cuts \cite{tang2021cutqc, peng2020simulating}). Therefore, in our example, reducing dependencies and reusing qubits is the most efficient solution in terms of virtualization overhead: we reduce the width of the circuit to $s=3$ with a virtualization budget of only $b=1$.
Note that the QR does not affect the execution of the VC on a distributed set of QPUs, since the QR only considers the reuse of qubits within the same fragment.

\subsection{Code Generator (CG)}
\label{sec:DT}

The final step of the \projecttitle{} compiler is generating the code in the form of circuits, which can be executed by the \projecttitle{} runtime (Fig.~\ref{fig:code-gen}).
To do so, we first extract each fragment as an individual circuit from the VC by collecting all operations on the respective qubit register.
In these extracted circuits, we insert placeholder gates at the qubits of the virtual gates.
The placeholder gates are parameterized gates, which can be instantiated with the actual gates that we need to insert to reconstruct the result (\S~\ref{sec:back:gate-virt}).
For the instantiation, the CG creates a parameter vector for each placeholder gate, which describes the gates to be inserted for the respective instance of the virtual gate.
Finally, the code-generator runs a set of standard circuit optimization passes to optimize the individual circuits.
This means the heavy optimization must be executed only once, reducing the just-in-time transpilation time before instance execution.

\section{The \projecttitle{} Runtime}
\label{sec:runtime}

\begin{figure*}[t] 
    \centering 
    \includegraphics[width=.74\textwidth]{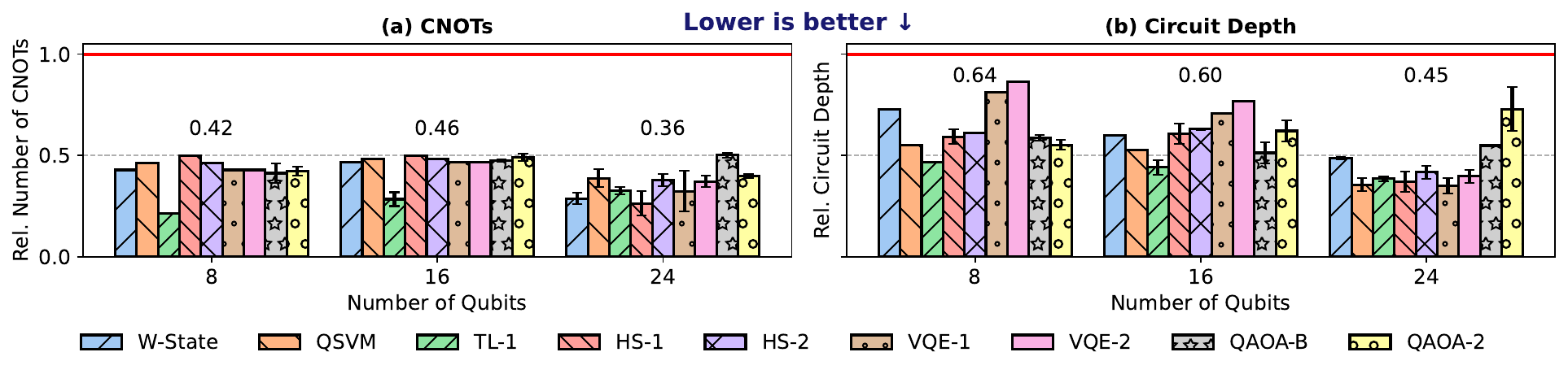} 
    \caption{Circuit Cutter (\S~\ref{sec:eval:circ_cutter}). {\em Impact of \projecttitle{}'s optimal circuit cutter compiler on the number of CNOTs and circuit depth.}}
    \label{plot:cut_depth_cnot}
    \vspace{-5mm}
\end{figure*}

\begin{figure*}[t] 
    \centering 
    \includegraphics[width=.74\textwidth]{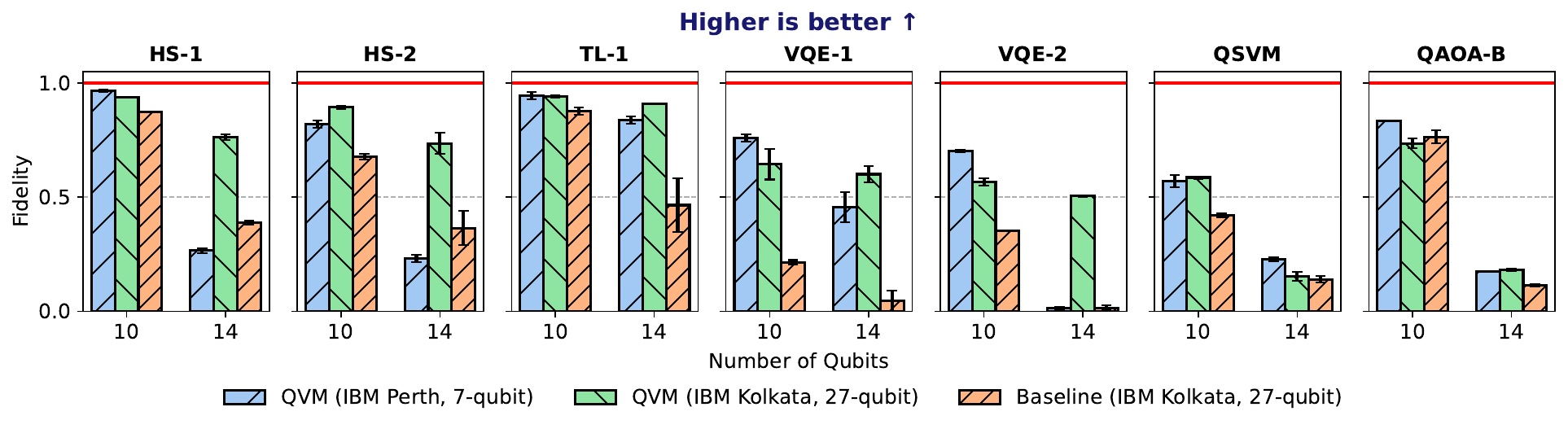} 
    \caption{Circuit Cutter (\S~\ref{sec:eval:circ_cutter}).  {\em Fidelity of running QVM with the circuit cutter compiler on IBM Perth and IBM Kolkata.}}
    \label{plot:cut_fid}
    \vspace{-5mm}
\end{figure*}

We next describe the \projecttitle{} runtime, a system for the scalable execution of virtualized circuits.
% The \projecttitle{} runtime is organized into two layers: the virtualizer and the scheduler, which we explain below. 

{\revision

\subsection{Workflow of the \projecttitle{} Runtime}
Fig. \ref{fig:runtime_overview} shows the workflow of the \projecttitle{} runtime.
As the first step, we pass optimized fragment circuits generated by the \projecttitle{} compiler to the \textit{virtualizer}.
Here, the \textit{instantiator} generates the instances of each circuit and passes them together with the fragments to the QPU manager. The \textit{QPU manager} then executes the fragments with each given instantiation on the distributed set of QPUs. The results are returned to the \textit{knitter} component of the virtualizer, where the final result is reconstructed through parallel classical post-processing.

\subsection{Virtualizer}
\label{sec:virtualizer}

The virtualizer implements the logic for executing virtual gates.
For this purpose, the virtualizer consists of two components, the \textit{instantiator} (Fig. \ref{fig:runtime_overview}, Step 1) and the \textit{knitter} (Fig. \ref{fig:runtime_overview}, Step 3).

\myparagraph{Instantiator}
The instantiator is responsible for creating instances of gates that must be inserted into the fragment.
For this purpose, the instantiator creates $6^{k_j}$ instances for each fragment $F_j$, where $k_j$ is the number of virtual gates that act on the qubits of $F_j$.
The instances are described as assignments to the parameterized gates and include every possible combination of the total $6^{k_j}$ combinations of each fragment. 
These assignments are essentially the tensor-product of the parameter vectors of the generated code for each fragment (Fig.~\ref{fig:code-gen}).

\myparagraph{Knitter}
The knitter takes the results of the probability distribution of all fragment instances and calculates the final result of the original circuit by applying the formulas for gate virtualization with highly parallel processing. (Section~\ref{sec:back:gate-virt}).
The results are given as vectors for each fragment $F_j$ with entries $\braket{O_j}_i$ with $i = 1, ..., 6^{k_j}$.
To knit the results, the knitter distributes the result vectors of each fragment to the available classical nodes, where each node is given the task of computing a part of the global $6^k$ instances. We determine this part by assigning an equal part of the global coefficient vector $\mathbf{C}$ to each node (Eq.~\ref{eq:coeff}).
In the example of Fig.~\ref{fig:runtime_overview}, we divide the coefficient vector $\mathbf{C}$ into two parts $\mathbf{C_1}$ and $\mathbf{C_2}$ and calculate the partial sum at each node over the instances corresponding to each coefficient. Finally, we calculate the sum of the two partial results to obtain the final result $\braket{O}$.
In this way, we are able to linearly scale the post-processing of the circuit virtualization with respect to the number of cores used.

}

\myparagraph{Extensibilty}
We implement a virtualizer for gate virtualization as presented in \cite{mitarai2021constructing}. However, the design of our virtualizer also allows us to implement other divide-and-conquer techniques effectively \cite{tang2021cutqc, Ayanzadeh2023frozenQbits}. Such techniques all follow the same workflow of our virtualizer and could, therefore, be easily integrated into the \projecttitle{} runtime.

\subsection{QPU Manager}\label{sec:runtime:scheduler}
The QPU manager is responsible for a scalable execution of the $6^{k_j}$ instances of each circuit fragment $F_j$ on a set of distributed QPUs, returning the result-vector for each fragment (Fig.~\ref{sec:runtime}, Step 2).
For this, the QPU manager receives an optimized circuit fragment (\S~\ref{sec:DT}) and all instance combinations generated by the instantiator. To execute a fragment, the QPU-manager does the following steps:

% \noindent
% \textbf{Maybe do architecture compilation}

\noindent
{\bf Step 1:} For each QPU $QPU_i$ with enough qubits to run the circuit, we transpile the circuit to, including mapping and routing on the physical qubits of $QPU_i$.
Note that this has to be done only once for each 
We then compute the estimated probability of success $esp(QPU_i)$ of executing the circuit on that QPU.
This is done by computing the cost of the errors induced by the gates and measurements on the assigned physical qubits, as described in \cite{nation2023suppressing}.

\noindent
{\bf Step 2:} We normalize the current job queue sizes of the QPUs by dividing the length of each job queue by the length of the maximum job queue. This yields a relative waiting time as $w(QPU_i) \in [0, 1]$, where a higher value means a longer waiting time for the job.

\noindent
{\bf Step 3:} We compute the score $s_i$ for each QPU, where $c_i  = \alpha \cdot (1 - w(QPU_i)) + \beta \cdot esp(QPU_i)$, and choose the QPU with the highest score to execute the corresponding fragment. The user can choose $\alpha$ and $\beta$ to provide either fast runtime or less noisy results.

\noindent
{\bf Step 4:} Finally, for each instance combination, we insert the instantiation into the transpiled fragment for the selected QPU, resulting in a total of $6^{k_j}$ circuits when $k_j$ gates act in the respective fragment $F_j$. These circuits are then sent to the QPU as a job for execution, and the results are returned to the virtualizer.

Our strategy of incorporating queue times and estimated probabilities of success into the QPU manager can be easily applied to the current cloud-centric quantum infrastructure, where our QPU manager would be a client for some quantum resources offered by cloud providers \cite{ravi2021adaptive}. Our solution is currently the most efficient, as there is little control over the cloud providers' internal queues.

\section{Evaluation}
\label{sec:eval}

\begin{figure*}[t] 
    \centering 
    \includegraphics[width=.74\textwidth]{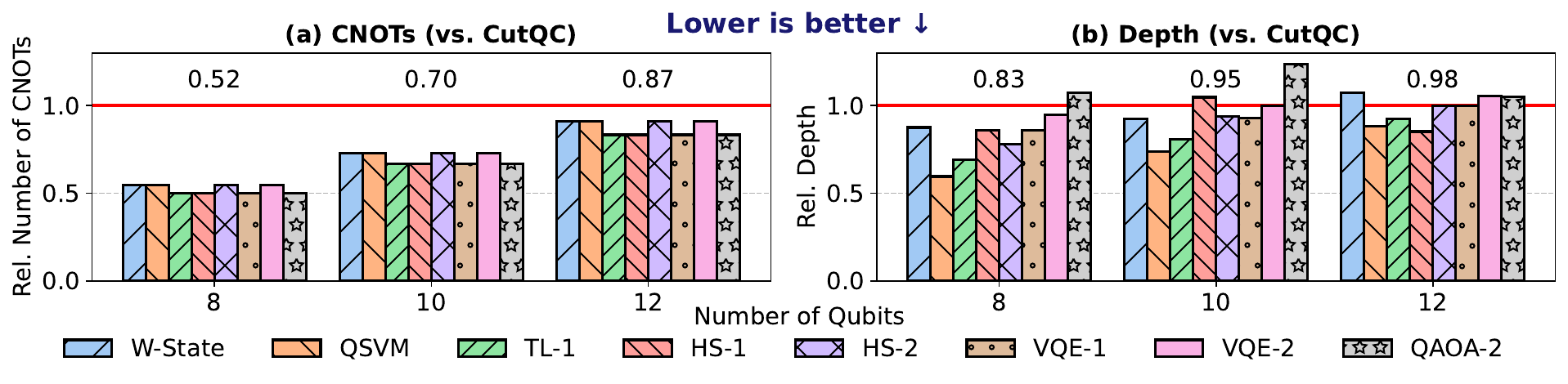} 
    \caption{\revision
    Circuit Cutter vs. CutQC (\S~\ref{sec:eval:circ_cutter}). {\em Relative number of CNOT gates and fragment depth compared to CutQC on IBM Kolkata.}}
    \label{plot:cutqc-stats}
    \vspace{-5mm}
\end{figure*}

\begin{figure*}[t] 
    \centering 
    \includegraphics[width=.74\textwidth]{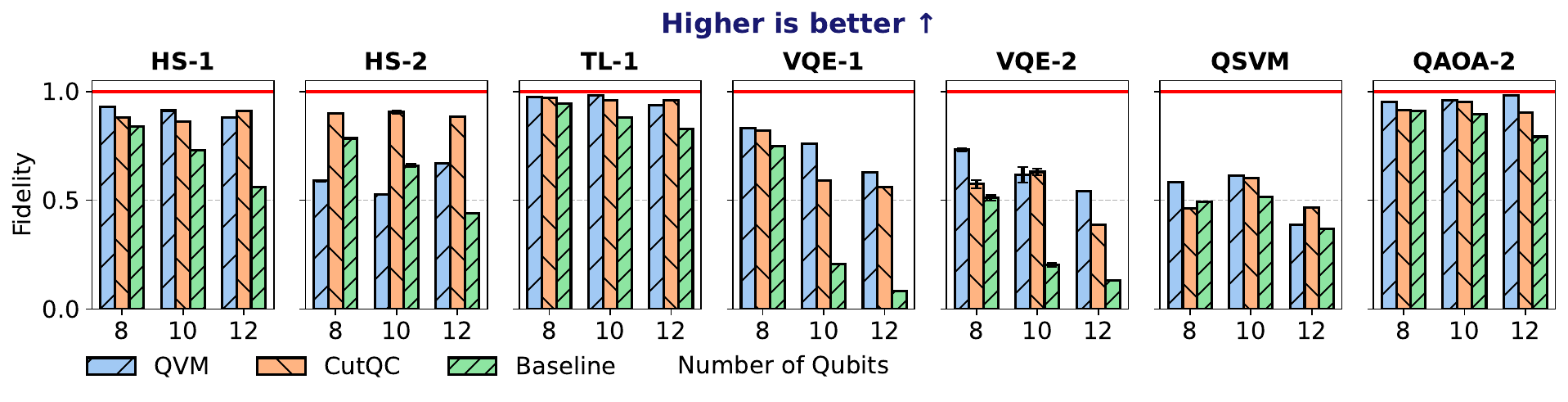} 
    \caption{\revision
    Circuit Cutter vs. CutQC (\S~\ref{sec:eval:circ_cutter}). {\em Fidelities of running \projecttitle{} vs. CutQC vs. Qiskit baseline on IBM Kolkata.}}
    \label{plot:cutqc-fid}
    \vspace{-5mm}
\end{figure*}

\myparagraph{Experimental Setup}
We conduct three types of experiments: (1) circuit transpilation with and without \projecttitle{}'s compiler to measure the circuit's properties post-compilation, (2) runs on real QPUs for measuring the circuit's fidelity, and (3) classical simulation of large circuits cut into fragments of different sizes.
For (2) we conduct our experiments on Falcon r5.11H QPUs, namely the 7-qubit IBM Perth and the 27-qubit IBMQ Kolkata.
%(Fig. \ref{fig:qpus}). 
For (1) and (3) we use the Qiskit Transpiler and Qiskit Aer \cite{qiskit-transpiler, qiskit-aer}, respectively, and run on our local classical machines.  For classical tasks, i.e., transpilation, post-processing (knitting), and simulation, we use a server with a 64-core AMD EPYC 7713P processor and 512 GB ECC memory.

\myparagraph{Framework and Configuration} We use the \textit{Qiskit} \cite{Qiskit} Python SDK version 0.41.0 for quantum circuits and simulations. We transpile any quantum circuit we run with the highest optimization level O3 and run with $20,000$ shots.
{\revision
To get a meaningful measurement of the fidelity or circuit properties on real QPUs, we run QVM only on a single QPU.
When we benchmark the performance of the \projecttitle{} runtime with simulators, we utilize every system core.
% Therefore, the evaluation is invariant to the scheduler's parameters $\alpha$ and $\beta$.
}

\myparagraph{Benchmarks}
We study \projecttitle{} on a set of circuits used in the state-of-the-art benchmark suits Supermarq \cite{tomesh2022supermarq}, MQT-Bench \cite{quetschlich2022mqt}, and QASM-Bench \cite{li2023qasmbench}. These circuits can be scaled both in the number of qubits and depth. Specifically, we study:
% \textup{GHZ},
\textup{W-State},
\textup{Bernstein Vazirani (BV)} , 
\textup{Quantum Support Vector Machine (QSVM)}, 
\textup{Hamiltonian Simulation (HS-$t$)}, 
\textup{Two Local Ansatz (TL-$n$)} with circular entanglement, 
\textup{Variational Quantum Eigensolver (VQE-$n$)} with a Real-Amplitudes ansatz of linear entanglement,
\textup{Approximate Optimization Algorithm (QAOA-$d$)} with regular graphs of degree $d\in\{2,3,4\}$ and barbell graphs (QAOA-B).
HS, VQE, and TL are scalable in their circuit layers $t$ or $n$, respectively.

\myparagraph{Metrics} We evaluate the following metrics.
\begin{itemize}
\item \textbf{Fidelity}: We use the \textit{Hellinger fidelity} to measure how close a noisy result is to the desired ground truth of a quantum circuit. The Hellinger fidelity is calculated as $\left(1 - H\left(P_{ideal}, P_{noisy}\right)^2\right)^2 \mapsto [0,1]$,
where $H$ is the Hellinger distance between two probability distributions \cite{hellinger1909neue}.
% {\revision We adopt this definition from Qiskit \cite{fidelity-qiskit}, as this definition is also used in the benchmark suite SupermarQ \cite{tomesh2022supermarq}.}

\item \textbf{Circuit Properties}: Number of \textit{CNOT} gates, \textit{depth} and the number of qubit \textit{dependencies}. When a VC contains more than one fragment, we use the fragment with the \textit{worst} property (i.e., maximal depth, dependecies, number of CNOTs)

\item \textbf{Execution Time}: The execution time of a VC in seconds.
% We compare the runtime of a classical circuit simulation with that of using \projecttitle{} on top of simulator(s).
% In our evaluation, the compilation and transpilation time is negligible compared to the runtime of quantum and classical computations of circuits.

{
\revision
\item \textbf{Estimated Success Probability}: We use the estimated success probability (ESP) metric to measure the estimated fidelity on larger quantum systems. We define the ESP as $\prod_i (1-e_i)$, where $e_i$ is the error of the $i$-th operation in the circuit \cite{mapomatic}. If a VC has multiple fragments, we report the minimum ESP.
}

\end{itemize}

\myparagraph{Baseline}
We use the Qiskit transpiler \cite{qiskit-transpiler} with O3 and CutQC as our baselines for circuit compilation and runtime evaluation \cite{circuit-knitting-toolbox, tang2021cutqc}. 
% {\revision For comparison with \projecttitle{}'s circuit cutter (\S~\ref{sec:eval:circ_cutter}) and runtime evaluation (\S~\ref{sec:eval:end-to-end-analysis}), we also compare against CutQC's wire-cutting \cite{tang2021cutqc, circuit-knitting-toolbox}}. 

\begin{figure*}[t] 
    \centering 
    \includegraphics[width=.77\textwidth]{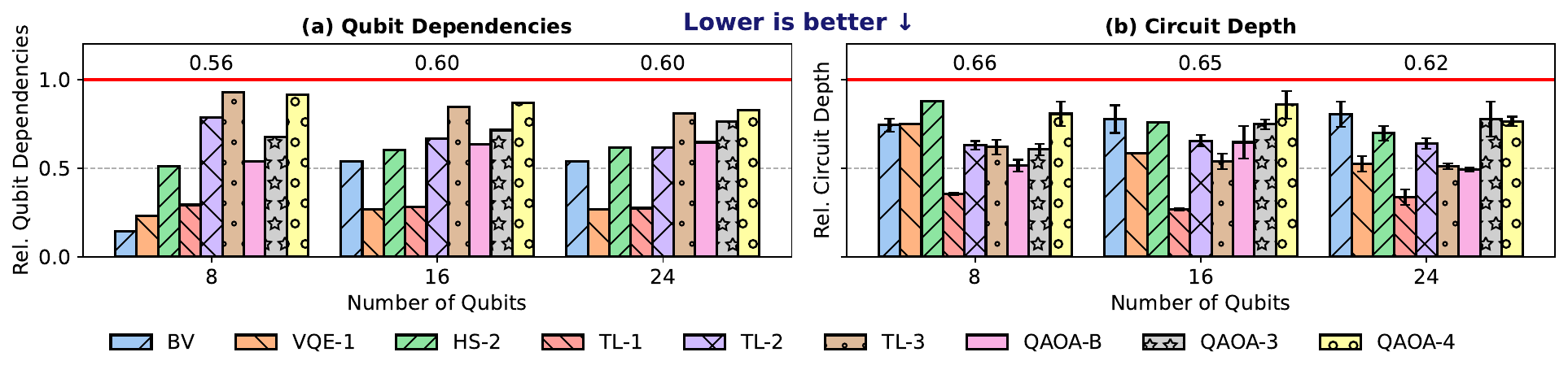} 
    \caption{Dependency Reducer (\S~\ref{sec:eval:dep_reducer}).  {\em Impact of the greedy qubit dependency reducer on (a) the number of qubit dependencies and (b) on the circuit depth of the transpiled circuit for IBM Kolkata. We use at most three virtual gates to compile the circuit.}}
    \label{plot:depmin_data}
    \vspace{-5mm}
\end{figure*}

\begin{figure*}[t] 
    \centering 
    \includegraphics[width=.77\textwidth]{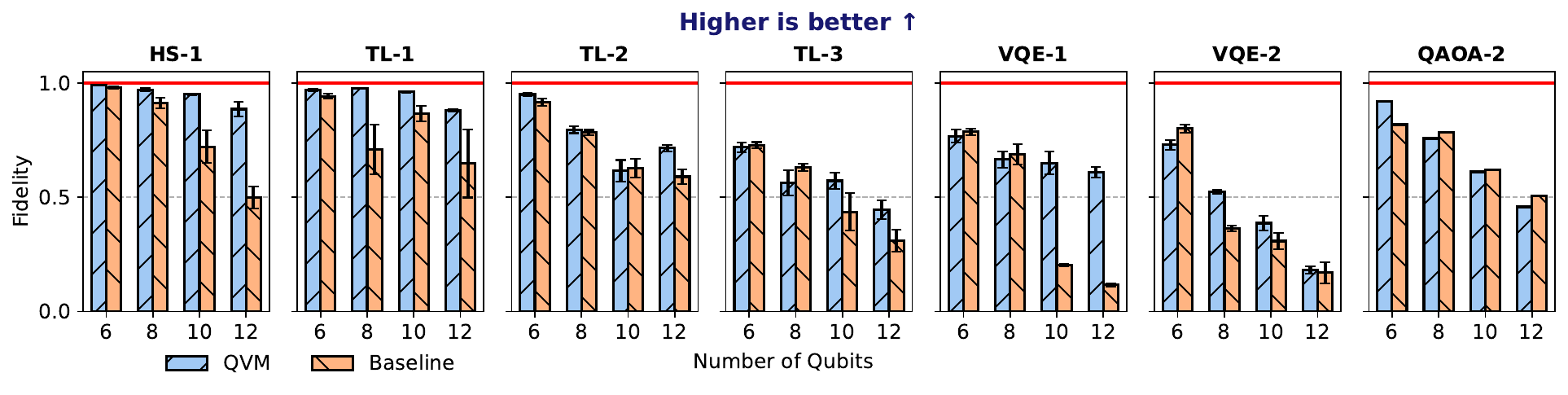} 
    \caption{Dependency Reducer (\S~\ref{sec:eval:dep_reducer}).  {\em Fidelity of the greedy dependency reducer using one virtual gate on IBM Kolkata.}}
    \label{plot:depmin_fid}
    \vspace{-5mm}
\end{figure*}

\subsection{Circuit Cutter}
\label{sec:eval:circ_cutter}

{\bf RQ1:} \textit{How well does \projecttitle{}'s circuit cutter allow scaling of circuits that can run on noisy QPUs with acceptable fidelity?}
We evaluate the impact of the circuit cutter on the CNOT count and depth of transpiled circuits and the fidelity of running virtual circuits using our optimal graph partitioning model.

\myparagraph{Impact on Number of CNOTs and Circuit Depth}
In Fig. \ref{plot:cut_depth_cnot}, we study the maximum number of CNOTs and circuit depths of the fragments after compilation with our circuit cutter with a maximum of three virtual gates.
Each virtual circuit is decomposed into fragments of a maximum of 13 qubits, and the fragments are transpiled for the 27-qubit IBMQ Kolkata QPU.
The results in Fig. \ref{plot:cut_depth_cnot} (a) show that the number of CNOTs decreases by 41\% on average. Fig. \ref{plot:cut_depth_cnot} (b) shows that the circuit depth decreases by 56\% on average.
This shows that it is possible to almost double the size of the circuits running with high fidelity on the given QPU since the number of CNOTs and circuit depth is approximately halved.

\myparagraph{Impact on Fidelity}
The impact of using \projecttitle{} on the fidelity of the execution is shown in Fig. \ref{plot:cut_fid}.
Here, the circuit cutter decomposes the circuits into fragments of maximally 7 qubits in order to theoretically fit the small 7-qubit IBM QPUs.
The fragments are run on both the 7-qubit IBM Perth and the 27-qubit IBM Kolkata QPUs, and compared to the baseline fidelity of running the circuits on IBM Kolkata.
We run the experiment for various benchmarks with sizes of 10 and 14 qubits.
We observe that the fidelity of running the circuit on IBM Kolkata improves the fidelity by 4.7$\times$ on average and up to 33.6$\times$. E.g. for the VQE-2 benchmark, the fidelity of the benchmark goes to zero, while \projecttitle{} can still create higher fidelities.
Compared to the baseline, running \projecttitle{} on the IBM Perth improves the fidelity by 2.1$\times$ on average and up to 10.6$\times$. Therefore we show that QVM can reliably simulate a larger QPU using smaller noisy QPUs while producing higher fidelity.
This is despite IBM Perth having a median of 2.3$\times$ higher readout and 1.2$\times$ higher CNOT error during our experiments.

{\revision

\myparagraph{Comparsion to CutQC}
In Fig. \ref{plot:cutqc-stats} and \ref{plot:cutqc-fid} we compare the circuit cutter of \projecttitle{} with CutQC.
We run the QVM and CutQC circuit cutters with the same configuration to generate circuit fragments of up to 7 qubits and compile and run the fragments on the IBM Kolkata QPU. We use several benchmarks of sizes of 8-12 qubits.
We find that, compared to CutQC, \projecttitle{} only produces 70\% of the CNOTs on average, since gate virtualization allows a reduction of the qubit connectivity significantly compared to CutQC (Fig. \ref{plot:cutqc-stats} (a)). \projecttitle{} achieves similar circuit depth reduction as CutQC as both can cut the circuits into significantly smaller fragments (Fig. \ref{plot:cutqc-stats} (b)).

A look at the fidelity benchmark (Fig. \ref{plot:cutqc-fid}) shows that CutQC and \projecttitle{} achieve similar fidelity and significantly outperform the Qiskit baseline, with \projecttitle{} achieving on average 1\% higher fidelity than CutQC.
We suspect the relatively small improvement despite the promising results in circuit properties is due to the noisy mid-circuit measurements with an error of $\geq 10^{-2}$, which we need to perform to virtualize gates. With less measurement noise, \projecttitle{} will likely perform similar or better to CutQC.

We conclude that both \projecttitle{} and CutQC, with their different techniques, are efficient in-circuit cutting and should ideally be used together in future work to take advantage of both methods with their respective benefits \cite{brandhofer2023optimal}, especially since we will likely be able to mitigate the mid-circuit measurement errors \cite{gupta2024probabilistic}.
}

\noindent\fbox{\parbox{\columnwidth}{
{\bf RQ1 takeaway:} With the circuit cutter, we reliably scale the size of circuits that can be run on noisy QPUs, up to 2$\times$, improving the overall fidelity 4.7$\times$ on average and up to 33.6$\times$ due to significant depth and CNOT gate reduction.
% \revision{The circuit cutter also provides benefits over CutQC in reducing the number of CNOT gates and achieves similar fidelities.}
}}

\subsection{Dependency Reducer}
\label{sec:eval:dep_reducer}

{\bf RQ2:} \textit{By how much does the dependency reducer (DR) decrease the number of dependencies within the circuit, improving the fidelity of running the circuit on noisy QPUs?}
For this experiment, we evaluate DR with a maximum of three virtual gates on several benchmarks with different circuit sizes on IBM Kolkata.

\myparagraph{Impact on Qubit Dependencies and Circuit Depth}
Fig.~\ref{plot:depmin_data} (a) shows the effect of DR on the number of qubit dependencies in the logical circuit, compared to the baseline of the circuit without DR. On average, the number of qubit dependencies decreases by 58\%. This shows that the DR can effectively resolve the dependencies between qubits, reducing noise propagation through the circuit.
As Fig.~\ref{plot:depmin_data} (b) shows, the depth of the circuits transpiled for IBM Kolkata decreases significantly by 64\% on average. This is due to the transpiler having fewer constraints on circuit mapping and routing after applying DR, resulting in a transpiled circuit with less depth.

\myparagraph{Impact on Fidelity}
We analyze the fidelity of our baseline and compared it to the DR in Fig.~\ref{plot:depmin_fid}, utilizing only one virtual gate. Our results indicate an average increase in fidelity of 36\% and up to 5.2$\times$. However, the noisy mid-circuit measurements needed for gate virtualization could limit the improvement in fidelity. These measurements typically induce significant noise, which affects the overall fidelity of virtual circuit execution \cite{yamamoto2022error, singh2023experimental}.

\noindent\fbox{\parbox{\columnwidth}{
{\bf RQ2 takeaway:} The DR decreases the dependencies between qubits by 58\% and circuit depth by 64\% using at most three virtual gates. This also leads to an average increase in fidelity by 36\% and up to 5.2$\times$, using only one virtual gate.
}}

\subsection{Tradeoffs with the Qubit Reuser}
\label{sec:eval:qr}

\begin{figure}[t] 
    \centering 
    \includegraphics[width=.7\columnwidth]{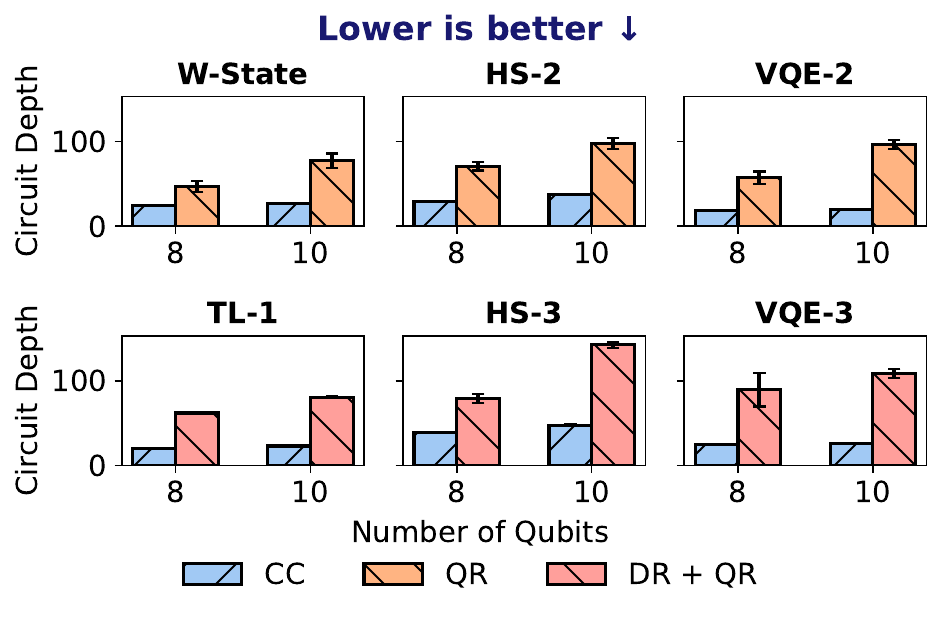} 
    \caption{Qubit Reuser (\S~\ref{sec:eval:qr}). {\em 
    Depths of compiled circuits with a maximal fragment size of 5 transpiled for IBM Perth. (a) Comparison of the circuit cutter vs. qubit reuser. (b) Comparison of the circuit cutter vs. dependency reducer and qubit reuser.}}
    \label{plot:eval:qr}
\end{figure}

\begin{figure*}[t] 
    \centering 
    \includegraphics[width=\textwidth]{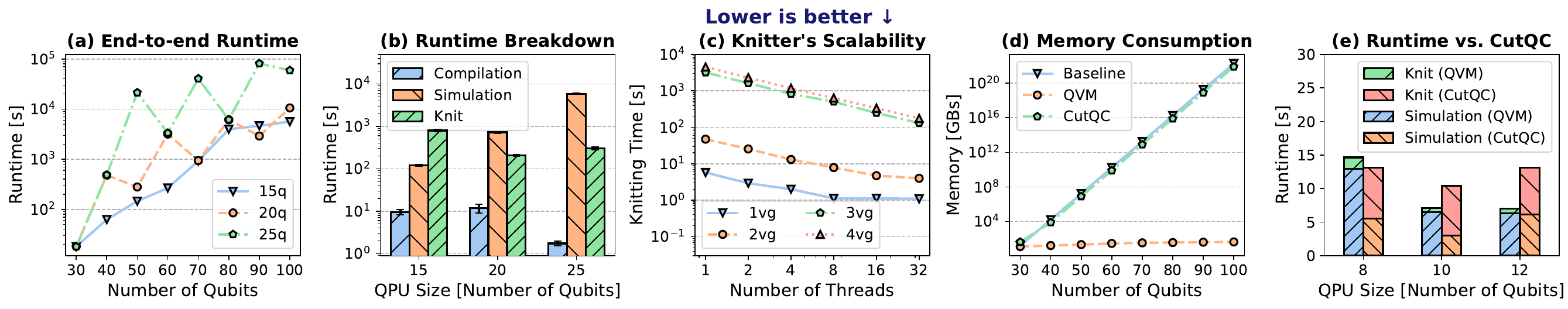} 
    \caption{\projecttitle{} end-to-end runtime analysis (\S~\ref{sec:eval:end-to-end-analysis}). {\em (a) End-to-end runtime of 30-100 qubits with different QPU sizes. (b) Runtime breakdown for 70 qubits with different QPU sizes. {\revision(c) Knit-time dependent on the number of parallel threads for different numbers of virtual gates (vg).} (d) Memory consumption for 30-100 qubits compared to the Baseline (statevector simulator) and Full Definition Query CutQC \cite{tang2021cutqc} with 20 qubits QPU size. {\revision(e) Runtime Comparison against CutQC for 20-qubit circuits.}}}
    \label{plot:runtime}
\end{figure*}

{\bf RQ3:} \textit{What is the effect of using the qubit reuser (QR) to reduce the width of the circuit fragments further?}
We show the trade-offs of using the CC alone against the DR and QR to reduce the width of circuits to run on small QPUs.
To show this tradeoff, we compile circuits with different optimizer configurations, such that each fragment's width is maximally five qubits. 
% We transpile the fragments to run on IBM Perth.

\myparagraph{Circuit-Cutter vs. Qubit-Reuse}
In Fig. \ref{plot:eval:qr} (top), we compare the effects of using either the CC or the QR to reduce the width of a virtual circuit on the circuit depth of the transpiled fragments.
Our results show that the CC compiles the circuits to only 37\% compared to QR on average. This is because the CC can break down the circuit into smaller fragments with reduced width while only incurring a maximum of two virtual gates. The QR, however, increases the depth of the circuit substantially while reusing qubits, which in turn will reduce overall fidelity.
So, there is a tradeoff between using gate virtualization to reduce the depth against using qubit reuse without overhead but with more depth.

\myparagraph{Combining Dependency Reducer and Qubit-Reuse}
In Fig. \ref{plot:eval:qr} (bottom), we show how the CC pass compares to the DR and QR passes to reduce circuit width. For this, we choose benchmarks where, without our DR, qubit-reuse would be impossible since every qubit depends on every other qubit in the circuit.
We apply the QR on the reduced-dependency circuit produced by the DR.
Like before, we aim to reduce the circuit width to five qubits.
The CC uses at most three, and the DR uses one virtual gate, with a 36$\times$ lower virtualization overhead.
Although using DR \& QR incurs a low overhead, it also leads to a significantly higher depth than CC. This means that the virtual circuit using DR \& QR has 3.2$\times$ more depth than the virtual circuit of CC, which could negatively impact the fidelity.

\noindent\fbox{\parbox{\columnwidth}{
{\bf RQ3 takeaway:} We find a trade-off between overhead and noise when using the CC or DR \& QR to reduce the width of quantum circuits.
While the CC produces circuits with smaller depths, combining DR and QR allows lower virtualization overhead.
}}

\subsection{\projecttitle{} End-to-end Runtime Analysis}
\label{sec:eval:end-to-end-analysis}

\begin{figure*}[t] 
    \centering 
    \includegraphics[width=.72\textwidth]{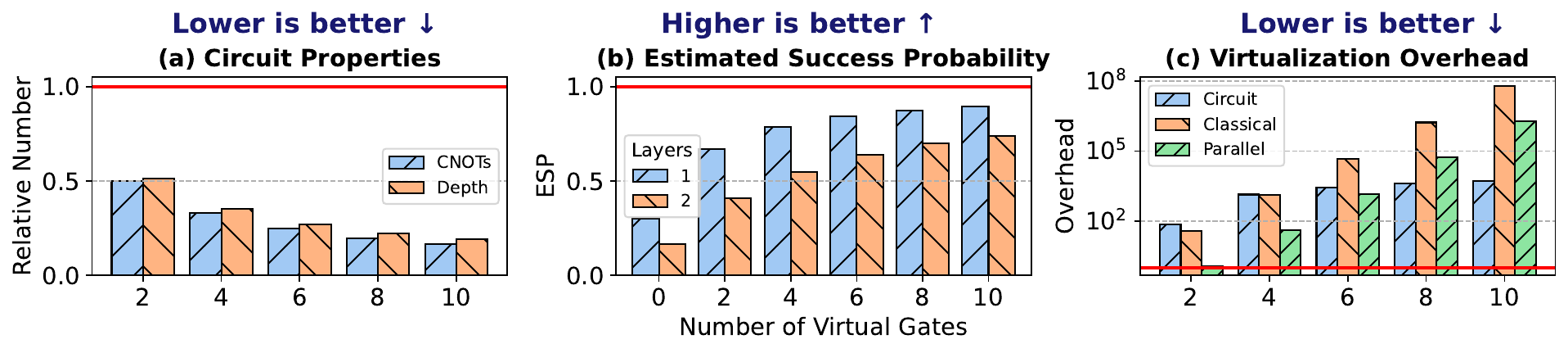} 
    \caption{\projecttitle{} at practical scale with 500 qubit \textit{VQE} circuits (\S~\ref{sec:eval:practical}). {\em (a) Relative number of CNOTs and circuit depth of the compiled VQE-2 benchmark. (b) Estimated success probability (ESP) with VQE-1 and VQE-2. (c) The overheads of circuit instances and classical postprocessing with and without parallel processing on 32 cores (linear speedup).}}
    \label{plot:large_scale}
    \vspace{-5mm}
\end{figure*}

{\bf RQ4:} \textit{{\revision How scalable is \projecttitle{}'s runtime and} how does \projecttitle{} compare to classical simulations without cutting \& knitting and CutQC?}
We study the HS-1 benchmark and use the circuit cutter (CC) to compile a VC for a QPU of up to $s$ qubits.
% a hypothetical simulated QPU of size $s$.
% We choose this simple circuit to keep the number of virtual gates required to cut the circuit proportional to the circuit and QPU size.

Fig. \ref{plot:runtime} (a) illustrates the end-to-end runtime needed to simulate \textit{HS-1} after cutting the circuit into fragments that fit QPUs of sizes $s\in\{15, 20, 25\}$. As the full circuit size increases, the runtime also increases, but the growth rate varies among fragment sizes. The smallest fragment size is the fastest, as the simulation overhead outweighs the knitting overhead, even if the circuit has 100 qubits and is cut with five virtual gates. 
This is evident in Fig. \ref{plot:runtime} (b) as well, which shows the runtime breakdown for simulating the 70-qubit \textit{HS-1}. As $s$ increases, there is a shift in the runtime from knitting to simulation time. The compilation time remains relatively constant. 

{\revision
Fig. \ref{plot:runtime} (c) shows the scalability of the knitter (\S~\ref{sec:virtualizer}) with its parallelism. We generate knit workloads for 1-4 virtual gates for the 70-qubit \textit{HS-1} benchmark and scale the number from 1 to 32 threads. We observe near-linear scalability with an increasing number of threads, allowing a speedup of up to 25.6$\times$ for 32 threads.
}

We show the memory required to simulate \textit{HS-1} with a chosen QPU size of 20 qubits in Fig. \ref{plot:runtime} (d). While the baselines, Qiskit Aer statevector, and CutQC with full definition query \cite{tang2021cutqc}, exhibit exponentially growing memory for linearly increasing circuit sizes, \projecttitle{} maintains a slightly increasing memory requirement by utilizing sparse quasi-probability distributions. In contrast, CutQC and simulations operate on tensors that need to cover the entire sample space.

{\revision
Finally, in Fig. \ref{plot:runtime} (e), we compare the runtimes of \projecttitle{} and CutQC. We are limited to comparing on small examples, due to CutQC's memory limitations. In particular, we perform 20-qubit circuits for \textit{HS-1} with simulated QPUs of 8-12 qubits.
We observe similar runtimes for the QPU size of 8 qubits, as \projecttitle{} spends more time to simulate a larger number of circuits due to the higher circuit cost \cite{peng2020simulating, tang2021cutqc}. However, with a QPU size of 10 and 12 qubits, \projecttitle{} clearly outperforms CutQC, as it achieves a significant acceleration in knitting due to its more efficient memory utilization.
}

\noindent\fbox{\parbox{\columnwidth}{
{\bf RQ4 takeaway:} \projecttitle{} enables simulating large circuits on classical simulators. It can handle circuit sizes of up to 100 qubits or five virtualized gates while maintaining acceptable runtime ($\sim$ 1.5 hours) and relatively very low memory consumption. {\revision \projecttitle{}'s knitter allows it to scale linearly through its high parallelization.}.}
}

\subsection{\projecttitle{} at Practical Scale}
\label{sec:eval:practical}
{\bf RQ5:} \textit{How does \projecttitle{} behave on a practical scale with circuits of hundreds of qubits?}
We would need hundreds to thousands of high-fidelity qubits to demonstrate quantum advantage. However, current QPUs with hundreds of qubits cannot reliably execute circuits with tens of qubits and higher depth.
To investigate how \projecttitle{} would behave on a practical scale, we evaluate the impact of \projecttitle{} on 500-qubit \textit{VQE} circuits on a heavy-hex lattice QPU with 883 physical qubits, which is the typical chip layout for current IBM QPUs \cite{ibmQuantum}.

\myparagraph{Impact on Number of CNOTs and Circuit Depth}
In Fig.~\ref{plot:large_scale} (a) we show the effects of the number of CNOTs and the circuit depth of the \textit{VQE-2} benchmark.
We see that using a budget of two virtual gates reduces the number of CNOTs and the circuit depth by 2$\times$, and using up to 10 virtual gates reduces the numbers by 6$\times$.
We see a diminishing improving impact on higher virtualization budgets.

\myparagraph{Impact on Estimated Success Probability}
Fig.~\ref{plot:large_scale}~(b) shows the estimated success probability (ESP) of the benchmarks \textit{VQE-1} and \textit{VQE-2}.
We find that the baseline without virtual gates achieves an ESP of only 30\% and 16\%, respectively, which leads to unusable results.
When using only two virtual gates, the ESP more than doubles and shows improvements, reaching 90\% and 74\% with 10 virtual gates.
This shows that with \projecttitle{}, we only need a handful of virtual gates to significantly improve the ESP, which could lead to usable results of quantum computation.

\myparagraph{Impact on Processing Overhead}
The virtualization costs incurred using virtual gates to improve circuit fidelity are shown in Fig.~\ref{plot:large_scale} (c).
The number of circuits that need to be instantiated and executed increases exponentially with a small number of virtual gates but then only starts to grow linearly with the number of fragments since we only instantiate as many circuits as correspond to the number of gates in the respective fragment (\S~\ref{sec:virtualizer}).
The classical post-processing overhead grows exponentially with $\mathcal{O}(6^k)$, meaning that adding two more virtual gates in the same configuration results in a runtime increase of 36$\times$.
Since the \projecttitle{} runtime provides an almost linear speedup (\S~\ref{sec:eval:end-to-end-analysis}), we can distribute the knitting across dozens of cores, which significantly mitigates this overhead for a small number of 4-6 virtual gates.
This is shown in Fig.~\ref{plot:large_scale} (c) as an example of (perfect) linear scaling in classical post-processing with 32 cores.
 
\noindent\fbox{\parbox{\columnwidth}{
{\bf RQ5 takeaway:} For large-scale algorithms, \projecttitle{} achieves high estimated success probability (ESP) while using only a handful of virtual gates for which our runtime can achieve significant speedups through parallelization. We therefore find a trade-off between fidelity and quantum-classical co-processing resources.}
}
\section{Related Work}

%\pramod{PLEASE TELL US A STORY: How the Quantum Computing is evolving...not just one bucket after the another. Make it more interesting.}
%\myparagraph{Circuit Cutting and Knitting}
%Circuit cutting and knitting, often referred to as \textit{circuit knitting} in short, is an umbrella term that covers a range of techniques aimed at diving quantum circuit into smaller fragments. Typically, the goal of using such techniques is to execute quantum circuits larger (wider) than the available quantum backends. We detail the prevalent circuit knitting techniques and compare them to gate virtualization  (\projecttitle{}).

\myparagraph{Quantum Transpilers and Error Mitigation}
%The development of better quantum circuit transpilation techniques to improve the fidelity of NISQ applications is an active area of research.
We can categorize quantum circuit transpilation techniques as (1) qubit mapping and routing \cite{gushu2019tackling, prakash2019noise, swamit2019not, chi2021time, molavi2022qubit, tannu2019ensemble, Zulehner2019anefficient, wille2019mapping, siraichi2018qubit, patel2022geyser, patel2021qraft}, (2) instruction/pulse scheduling \cite{das2021adapt, tripath2022suppression, zlokapa2020deep, smoth2022timestitch, murali2020software, yunong2019optimized, gokhale2020optimized} and (3) gate optimization/decomposition \cite{yunong2019optimized, patel2022geyser, litteken2023dancing, das2023the, xu2023synthesizing, patel2022quest}. Finally, there is work on post-execution processing, readout improvement, and error correction \cite{maurya2023scaling, bravyi2021mitigating, maciejewski2020mitigation, patel2020veritas, tannu2022hammer, das2021jigsaw, patel2020disq, dangwal2023varsaw, tannu2019mitigating}. These proposals are orthogonal to our work and can be integrated into \projecttitle{}. This is especially the case for measurement error mitigation, which can help to improve the fidelity of the mid-circuit measurements during execution \cite{yamamoto2022error, singh2023experimental}.

% However, these works are not able to optimize quantum circuits in the way that \projecttitle{} does, i.e., reduce their width and/or qubit dependencies.
%map and route a circuit's qubits across a set of QPUs, nor do they use a circuit-cutting technique to further improve qubit routing, i.e. reduce the number of swap operations.

%\nate{We might need something about QPD here}
\myparagraph{Circuit Cutting and Knitting} %\myparagraph{Wire cutting}
%\myparagraph{Circuit Cutting and Knitting}
Circuit cutting \& knitting is the process of breaking down a large quantum circuit into smaller sub-circuits that can be executed separately, then synthesizing the results to obtain the result of the original circuit. Circuit cutting can be divided into gate virtualization (\S~\ref{sec:back:gate-virt}) and \textit{wire cutting} \cite{peng2020simulating, Bravyi2016trading, tang2021cutqc, brenner2023optimal, ufrecht2023cutting}. % Wire cutting divides the circuit based on the evolution of qubit wires over time, enabling independent simulation of qubit wires. 
While wire cutting optimizes circuits for small QPUs with reduced noise, it is limited in reducing qubit dependencies. Our work proposes a generic architecture for gate virtualization. Furthermore, wire cutting can be simulated using gate virtualization \cite{brenner2023optimal, circuit-knitting-toolbox}. %Still, a hybrid approach combining wire cutting and gate virtualization could leverage the advantages of both techniques.
%Finally, although Qiskit's Circuit Knitting Toolbox \cite{circuit-knitting-toolbox} implements both techniques, finding the cut locations remains a manual process, lacking an automated way to determine optimal cut locations.

\myparagraph{Qubit Reuse}
Qubit reuse can be classified into two categories, namely ancilla reuse using \textit{uncomputation}  \cite{Bennett1973logical}, and reuse through \textit{dynamic circuits} \cite{ibm-roadmap, dynamic-circuits}.
%Uncomputation refers to the process of re-applying the operations of a qubit in the reverse order to bring the qubit back to its initial state. 
Work such as \cite{bichsel2020silq, paradis2021unqomp, ding2020square} utilize uncomputation to reclaim ancilla qubits. In contrast, work such as \cite{paler2016wire, hua2023caqr, sadeghi2022quantum, decross2022qubitreuse} exploit the newly supported dynamic circuits with mid-circuit measurements and mid-circuit reset operations to reuse qubits. 
%Dynamic circuits \cite{ibm-roadmap, dynamic-circuits} allow for the creation, deletion, and modification of gates and qubits within the circuit based on the intermediate measurements and computational results obtained during execution. 
%Dynamic circuits \cite{ibm-roadmap, dynamic-circuits} allow for the creation, deletion, and modification of gates and qubits within the circuit based on the intermediate measurements and computational results obtained during execution. Papers including \cite{paler2016wire, hua2023caqr, sadeghi2022quantum, decross2022qubitreuse} combine mid-circuit measurements with mid-circuit reset operations to allow the qubits to be reused in subsequent parts of the circuit, enabling more efficient use of limited qubit resources.
However, applying these techniques on densely connected circuits can be impractical due to the large number of qubit dependencies 
\cite{tomesh2022supermarq, hua2023caqr}.
% \cite{Ayanzadeh2023frozenQbits}
% For instance, specific quantum algorithms are represented as densely connected or power-law graphs with high qubit communication \cite{Ayanzadeh2023frozenQbits, tomesh2022supermarq}. 
% For tracking qubit dependencies, the work in \cite{hua2023caqr} states an overhead of $O(n^2)$, where $n$ is the number of gates. 
By first applying \projecttitle{}'s DR pass (\S~\ref{sec:cdr}), qubit reuse can be practically applied with enhanced efficiency.

\myparagraph{Application-specific Optimizations}
Application-specific circuit optimizations go beyond generic strategies and target the unique characteristics of a particular algorithm or circuit structure in order to improve fidelity~\cite{alam2020circuit, lao20222qan, gokhale2019partial, gokhale2019minimizing, li2022paulihedral, stein2022eqc, resch2021accelerating, hao2023enabling, tuysuz2023classical, Ayanzadeh2023frozenQbits}. Our work tries to build a generic and extensible framework to incorporate different application-specific optimizations in the context of gate virtualization. 

%One such technique is FrozenQubits \cite{Ayanzadeh2023frozenQbits}, specialized for QAOA circuits, which leverages the power-law distribution of QAOA graphs.  However, FrozenQubits' limitation is its specificity to QAOA circuits, especially those following a power-law node distribution. The other approaches adopt specialized techniques that target their respective applications but lack generality. In contrast, \projecttitle{} is a generic approach.

%The rest of the approaches are diverse technique-wise, i.e., they do not necessarily follow similar methodologies for transforming the input quantum circuits. Although they are sophisticated and effective for their respective types of applications, their key limitation is the lack of generality. In contrast, \projecttitle{} is generic and applicable to any type of application.

%This partitions the problem into smaller sub-problems with reduced CNOT operations. Other approaches that are application-specific, e.g., for variational quantum algorithms, include \cite{alam2020circuit, lao20222qan, gokhale2019partial, gokhale2019minimizing, li2022paulihedral, stein2022eqc, resch2021accelerating}. 

\myparagraph{Quantum Cloud Computing} This area addresses quantum circuit multi-programming \cite{das2019a, liu2021qucloud, niu2023enabling, ohkura2022simultaneous}, quantum resource management/scheduling \cite{ravi2021adaptive, ravi2021quantum, weder2021automated}, and  quantum serverless \cite{nguyen2022qfaas, garcia-alonso2022quantum}. Our work is complimentary to these proposals, \projecttitle{} proposes a scalable infrastructure for supporting gate virtualization optimizations, which can be incorporated by quantum cloud environments. 

\section{Conclusion}
We introduce the Quantum Virtual Machine (\projecttitle{}), a generic system for scalable, high-fidelity execution of large circuits on noisy and small QPUs by leveraging gate virtualization.  \projecttitle{} extends the quantum circuit abstraction with the \textit{virtual circuit IR}, which forms the foundation for the \projecttitle{} Compiler---a modular compiler infrastructure for implementing a series of optimization passes to generate smaller, optimized fragments. These fragments are virtualized and executed using our \projecttitle{} Runtime---a distributed and scalable system to execute and post-process the instantiated circuit fragments in a highly parallel manner on a distributed set of QPUs. Our evaluation on IBM's 7- and 27-qubit QPUs of \projecttitle{} demonstrates practical scaling of circuits with sizes up to double the QPU capacity while significantly improving fidelity.% Additionally, \projecttitle{} effectively reduces circuit depths using the compiler and achieves remarkable scalability up to 100 qubits in circuit simulation, outperforming classical simulators. 

% \myparagraph{Artifact}
% \projecttitle{} is publicly available for artifact evaluation.

\section{Acknowledgments}
We thank Karl Jansen and Stefan Kühn from the Center for Quantum Technology and Applications (CQTA)- Zeuthen for supporting this work by providing access to IBM quantum resources. We thank Martin Ruefenacht for his valuable contributions during his employment at the Leibniz Supercomputing Center (LRZ). We also thank Ahmed Darwish for his contributions to this work.

We acknowledge the use of IBM Quantum services for this work. The views expressed are those of the authors, and do not reflect the official policy or position of IBM or the IBM Quantum team.
Funded by the Bavarian State Ministry of Science and the Arts as part of the Munich Quantum Valley (MQV).

\bibliographystyle{plain}
\bibliography{refs}

\end{document}